\documentclass[cpp,fleqn]{w-art}
\usepackage{times}
\usepackage{w-thm}
\usepackage{lineno,hyperref}
\usepackage[]{graphicx}
\usepackage{amsmath}
\usepackage{amssymb}
\usepackage{mathtools}
\usepackage{braket}

\DeclareMathOperator*{\SumInt}{%
\mathchoice%
  {\ooalign{$\displaystyle\sum$\cr\hidewidth$\displaystyle\int$\hidewidth\cr}}
  {\ooalign{\raisebox{.14\height}{\scalebox{.7}{$\textstyle\sum$}}\cr\hidewidth$\textstyle\int$\hidewidth\cr}}
  {\ooalign{\raisebox{.2\height}{\scalebox{.6}{$\scriptstyle\sum$}}\cr$\scriptstyle\int$\cr}}
  {\ooalign{\raisebox{.2\height}{\scalebox{.6}{$\scriptstyle\sum$}}\cr$\scriptstyle\int$\cr}}
}

\newcommand{\op}[1]{\ensuremath{\hat{#1}}}
\newcommand{\ladderdown}{\ensuremath{\op{a}^{\vphantom{\dagger}}}}
\newcommand{\ladderup}{\ensuremath{\op{a}^\dagger}}
\newcommand{\annihilop}{\ladderdown}
\newcommand{\creationop}{\ladderup}

\begin{document}
\DOIsuffix{theDOIsuffix}
\Volume{42}
\Issue{1}
\Month{01}
\Year{2003}
\pagespan{1}{}
\Receiveddate{September 2017}
\Reviseddate{--}
\Accepteddate{--}
\Dateposted{--}



\title[\textit{Ab Initio} results for the...]{\textit{Ab Initio} results for the Static Structure Factor\\ of the Warm Dense Electron Gas}


\author[Tobias Dornheim]{Tobias Dornheim\footnote{Corresponding
     author: e-mail: {\sf dornheim@theo-physik.uni-kiel.de}}\inst{1}} \address[\inst{1}]{Institut f\"ur Theoretische Physik und Astrophysik, Christian-Albrechts-Universit\"{a}t zu Kiel, D-24098 Kiel, Germany}
\author[Simon Groth]{Simon Groth\inst{1}}

\author[Michael Bonitz]{Michael Bonitz\inst{1}}
\begin{abstract}
The uniform electron gas at finite temperature is of high current interest for warm dense matter research. The complicated interplay of quantum degeneracy and Coulomb coupling effects is fully contained in the pair distribution function or, equivalently, the static strucutre factor.
By combining exact quantum Monte Carlo results for large wave vectors with the long-range behavior from the Singwi-Tosi-Land-Sj\"olander approximation, we are able to obtain highly accurate data for the static structure factor over the entire $k$-range.
This allows us to gauge the accuracy of previous approximations and discuss their respective shortcomings. Further, our new data will serve as valuable input for the computation of other quantities.
\end{abstract}
\maketitle                   





\section{Introduction}
Over recent years, there has emerged a growing interest in warm dense matter (WDM) -- an exotic state where strong electronic excitations are realized at solid state densities~\cite{wdm_book}. In addition to astrophysical applications such as planet interiors~\cite{knudson,militzer} and white dwarf atmospheres, such extreme conditions are now routinely created in the lab, e.g., in experiments with laser excited solids~\cite{ernst} or inertial confinement fusion~\cite{nora,schmit,hurricane3}. Despite this remarkable experimental progress, a rigorous theoretical description remains notoriously difficult due to the simultaneous presence of three physical effects: (1) strong electronic excitations, (2) Coulomb coupling effects, and (3) fermionic exchange. This is typically expressed by two parameters being of the order of unity: the degeneracy temperature $\theta=k_\textnormal{B}T/E_\textnormal{F}$ (with $E_\text{F}=k_\text{F}^2/2$ and $k_\text{F}=(9\pi/4)^{1/3} / r_s$ being the Fermi energy and wave vector, respectively) and the Brueckner (coupling) parameter $r_s=\overline{r}/a_\textnormal{B}$ with $\overline{r}$ and $a_\textnormal{B}$ being the mean interparticle distance and Bohr radius, respectively.

Of particular importance is the calculation of the thermodynamic properties of the uniform electron gas (UEG), which is comprised of Coulomb interacting electrons in a homogeneous neutralizing background. However, this has turned out to be surprisingly difficult. The extension of Quantum Monte Carlo (QMC) methods, which have been employed to obtain very accurate data in the ground state already three decades ago~\cite{gs1,gs2}, to finite temperature is severely limited by the fermion sign problem~\cite{loh,troyer}. It was only recently that the combination of two novel methods (configuration path integral Monte Carlo [CPIMC]~\cite{tim_cpp15,tim3} and permutation blocking path integral Monte Carlo [PB-PIMC]~\cite{dornheim,dornheim2}) that are available at complementary parameter ranges allowed to conduct the first unbiased simulation of the UEG. At first, these efforts were limited to a finite number of electrons $N$ in a finite simulation cell of volume $V$, see Refs.~\cite{groth,dornheim3}.
In practice, however, one is interested in the thermodynamic limit, which is given by the limit of an infinite number of particles at fixed density (or, equivalently, fixed $r_s$).
This was realized by combining QMC data, which exactly incorporates all short-range exchange-correlation effects, but cannot capture the long-range effects due to the finite simulation cell, with the linear response theory, which is exact precisely in this limit~\cite{drummond,chiesa,dornheim_prl,dornheim_pop}.
The resulting accurate data for the UEG in the thermodynamic limit have subsequently been used to construct a complete parametrization of the exchange-correlation free energy with respect to temperature, density and spin-polarization over the entire warm dense matter regime~\cite{groth_prl,groth2}.

In this work, we further explore this strategy to investigate the static structure factor (SSF), $S(k)$, of the UEG at warm dense matter conditions. In particular, we construct cubic basis splines to combine the SSF from the Singwi-Tosi-Land-Sj\"olander theory (STLS)~\cite{stls_original,stls,stls2}, which is exact in the limit of small wave vectors ($k\to0$)~\cite{kugler2}, with the exact QMC data elsewhere. These new extensive data for $S(k)$ are subsequently compared both to the random phase approximation (RPA)~\cite{rpa_original} and the full STLS results themselves over two orders of magnitude of the coupling parameter $r_s$ and for three different temperatures.
This allows us to gauge the performance of the dielectric approximations and to show when they break down.

\section{Theory}

\subsection{The Uniform Electron Gas}
The uniform electron gas is defined as an infinite system of Coulomb interacting electrons in a uniform positive background ensuring charge neutrality.
Since QMC simulations are only posslibe in a finite simulation cell with box length $L$ and volume $V=L^3$, we employ periodic boundary conditions and the standard Ewald summation to take into account the interactions of the electrons with the infinite array of periodic images. Since PB-PIMC and CPIMC are formulated in coordinate space and momentum space, respectively, both representations of the UEG Hamiltionian are given. We assume Hartree atomic units throughout this work.
\subsection{Coordinate representation of the Hamiltonian\label{cor}}
Following Refs.~\cite{fraser,dornheim2}, we express the Hamiltonian for $N=N_\uparrow+N_\downarrow$ unpolarized ($N_\uparrow=N_\downarrow$) electrons in coordinate space as 
 \begin{eqnarray}
  \hat{H} = - \frac{1}{2}\sum_{i=1}^N \nabla^2_i +  \frac{1}{2}\sum_{i=1}^N\sum_{j\ne i}^N \Psi( \mathbf{r}_i, \mathbf{r}_j) + \frac{ {N} }{2}\xi_\textnormal{M} \; ,
  \label{Hcoord}
 \end{eqnarray}
with the Madelung constant $\xi_\textnormal{M}$ and the periodic Ewald pair interaction
\begin{eqnarray}
\Psi(\mathbf{r}, \mathbf{s} ) = \frac{1}{V} \sum_{ \mathbf{G} \ne 0 } \frac{ e^{-\pi^2\mathbf{G}^2/\kappa^2} e^{2\pi i \mathbf{G}(\mathbf{r}-\mathbf{s})} }{ \pi\mathbf{G}^2}
  \label{pair}  - \frac{\pi}{\kappa^2 V} + \sum_\mathbf{R} \frac{ \textnormal{erfc}( \kappa | \mathbf{r}-\mathbf{s} + \mathbf{R} | ) }{ |\mathbf{r}-\mathbf{s}+\mathbf{R} | } \ .
\end{eqnarray}
Here $\mathbf{R}=\mathbf{n}_1L$ and $\mathbf{G}=\mathbf{n}_2/L$ denote the real and reciprocal space lattice vectors, respectively, with $\mathbf{n}_1$ and $\mathbf{n}_2$ three-component vectors of integers, and $\kappa$ denotes the (freely adjustable) Ewald parameter.

 \subsection{Hamiltonian in second quantization\label{sec:second_quant}}
In second quantization with respect to spin-orbitals of plane waves,
\begin{eqnarray}
\langle \mathbf{r} \sigma \;|\mathbf{k}_i\sigma_i\rangle = \frac{1}{L^{3/2}} e^{i\mathbf{k}_i \cdot \mathbf{r}}\delta_{\sigma,\sigma_i}\ ,
\end{eqnarray}
with $\mathbf{k}_i=\frac{2\pi}{L}\mathbf{m}_i$, $\mathbf{m}_i\in \mathbb{Z}^3$ and $\sigma_i\in\{\uparrow,\downarrow\}$, 
the Hamiltonian, Eq.~(\ref{Hcoord}), is expressed as
\begin{eqnarray}\label{eq:h} 
& \op{H} =
\frac{1}{2}\sum_i \mathbf{k}_i^2 \creationop_{i}\annihilop_{i} +
\smashoperator{\sum_{\substack{i<j,k<l \\ i\neq k,j\neq l}}} 
w^-_{ijkl}\creationop_{i}\creationop_{j} \annihilop_{l} \annihilop_{k} + \frac{N}{2}\xi_M.
\end{eqnarray}
Here, the antisymmetrized two-electron integrals are defined as $w^-_{ijkl} =w_{ijkl}-w_{ijlk}$, with
\begin{align} 
\; w_{ijkl}=\frac{4\pi e^2}{L^3 (\mathbf{k}_{i} - \mathbf{k}_{k})^2}\delta_{\mathbf{k}_i+\mathbf{k}_j, \mathbf{k}_k + \mathbf{k}_l}\delta_{\sigma_i,\sigma_k}\delta_{\sigma_j,\sigma_l}\ ,
\label{eq:two_ints}
\end{align}
and the Kronecker deltas ensure both momentum and spin conservation. The first (second) term in the Hamiltonian, Eq.~(\ref{eq:h}), describes the kinetic (interaction) energy.
As usual, the operator 
$\creationop_{i}$  ($ \annihilop_{i}$) 
creates (annihilates) a particle in the (spin-) orbital $|\mathbf{k}_i\sigma_i\rangle$.

\subsection{Quantum Monte Carlo Simulations}
The task at hand to be solved using QMC methods is the calculation of canonic expectation values (temperature $T$, volume $V$, and particle number $N$ are fixed), that follow from the canonic partition function
\begin{eqnarray}\label{eq:basic_Z}
Z = \textnormal{Tr}\op\rho\ ,
\end{eqnarray}
with $\op\rho = e^{-\beta\op H}$ being the canonic density operator and the inverse temperature $\beta=1/k_{\textnormal{B}}T$. In particular, the thermodynamic expectation value of an arbitrary observable $\op A$ can be written as
\begin{eqnarray}\label{eq:thex}
\braket{\op A} = \frac{1}{Z} \textnormal{Tr} \op\rho \op A\ .
\end{eqnarray}
The underlying idea of both the CPIMC and the PB-PIMC method is to find a representation of the partition function Eq.~(\ref{eq:basic_Z}) of the form
\begin{eqnarray}\label{eq:MC_Z}
Z = \SumInt_\mathbf{C} W(\mathbf{C})\ ,
\end{eqnarray}
i.e., as a sum or integral over some, in general, high-dimensional variable $\mathbf{C}$, which is denoted as a configuration. The function $W(\mathbf{C})$ is the corresponding ''configuration weight'', which must be of a form that can be readily evaluated. The latter specification is not trivial as, for interacting electrons, the matrix elements of the density operator are not known when quantum effects are not negligible. 
Once a representation of the form of Eq.~(\ref{eq:MC_Z}) is found, the thermodynamic expectation value, Eq.~(\ref{eq:thex}), becomes
\begin{eqnarray}\label{eq:MC_A}
\braket{\op A} = \frac{1}{Z} \SumInt_\mathbf{C} W(\mathbf{C}) A(\mathbf{C})\ ,
\end{eqnarray}
with $A(\mathbf{C})$ being the so-called Monte Carlo estimator.
In practice, we use the Metropolis algorithm~\cite{metropolis} to generate a set of $N_\textnormal{MC}$ random configurations $\{\mathbf{C}_1,\dots,\mathbf{C}_{N_\textnormal{MC}}\}$ that are distributed according to the probability $P(\mathbf{C}) = W(\mathbf{C})/Z$, which is possible without explicit knowledge of the normalization $Z$. The Monte Carlo estimate for the thermodynamic expectation value from Eq.~(\ref{eq:MC_A}) is then given by
\begin{eqnarray}\label{eq:asdf}
\braket{\op A} \approx \braket{\op A}_\textnormal{MC} = \frac{1}{N_\textnormal{MC}}\sum_{i=1}^{N_\textnormal{MC}}A(\mathbf{C}_i)\ ,
\end{eqnarray}
which, in the limit of infinitely many random samples, $N_\textnormal{MC}\to\infty$, becomes exact
\begin{eqnarray}
\braket{\op A} = \lim_{N_\textnormal{MC}\to\infty} \braket{\op A}_\textnormal{MC}\ ,
\end{eqnarray}
where the Monte Carlo error for any finite number of samples is given by
\begin{eqnarray}
\Delta A = \left( \frac{\braket{A^2} - \braket{A}^2}{N_\textnormal{MC}} \right)^{1/2}\ .
\end{eqnarray}
Since the Monte Carlo estimates are exact within this statistical uncertainty, which is known accurately as well and can be made arbitrarily small by generating more random configurations, QMC simulations are often denoted as ''quasi-exact''.

Unfortunately, quantum Monte Carlo simulations of electrons are not so straightforward as we shall briefly illustrate in the following.
Due to the antisymmetry of the many-fermion wave function under exchange, the weight function $W$ in Eq.~(\ref{eq:MC_Z}) can be both positive or negative. This, in turn, means that $P(\mathbf{C}) = W(\mathbf{C}) / Z $ cannot be interpreted as a probability, which must be strictly positive.  In order to still be able to use the Metropolis algorithm, we switch to a modified configuration space (indicated by the ''prime'' symbols) where the configurations are sampled according to the modulus weights
\begin{eqnarray}
Z' = \SumInt_\mathbf{C} |W(\mathbf{C})|\ ,
\end{eqnarray}
and the definition of the modified expectation value
\begin{eqnarray}
\braket{\op A}' = \frac{1}{Z'} \SumInt_\mathbf{C} A(\mathbf{C}) |W(\mathbf{C})|\ .
\end{eqnarray}
The unbiased fermionic expectation value (\ref{eq:MC_A}) is then given by
\begin{eqnarray}\label{eq:FMC}
\braket{\op A} = \frac{ \braket{\op A \op S}' }{ \braket{\op S}' }\ ,
\end{eqnarray}
where $S(\mathbf{C})=W(\mathbf{C})/|W(\mathbf{C})|$ is the so-called sign and, thus, $S=\braket{\op S}'$ the ''average sign'' of the corresponding Monte Carlo simulation.
It is important to note that the statistical uncertainty of the Monte Carlo estimation according to Eq.~(\ref{eq:FMC}) is (in leading order) inversely proportional to $S$,
\begin{eqnarray}\label{eq:a}
\frac{ \Delta A }{A} \sim \frac{1}{ \braket{\op S}' \sqrt{N_\textnormal{MC}} }\ ,
\end{eqnarray}
while the average sign itself exponentially decreases both with inverse temperature and system size,
\begin{eqnarray}\label{eq:b}
\braket{\op S}' \sim e^{-\beta N (f-f')}\ ,
\end{eqnarray}
where $f$ denotes the free energy per particle.
Inserting Eq.~(\ref{eq:b}) into (\ref{eq:a}) leads to 
\begin{eqnarray}\label{eq:FSC}
\frac{{\Delta A}}{A} \sim \frac{e^{\beta N (f-f')}}{\sqrt{N_\textnormal{MC}}}\ .
\end{eqnarray}
Evidently, the statistical uncertainty exponentially increases both with system size and inverse temperature, which can only be compensated by increasing the number of Monte Carlo samples, thereby decreasing $\Delta A$ with the inverse square root of ${N_\textnormal{MC}}$. This is the notorious fermion sign problem~\cite{loh,troyer,dornheim_pop}, which has, for a long time, prevented \textit{ab initio} path integral Monte Carlo (PIMC, see Ref.~\cite{cep} for a review) simulations of electrons in the warm dense matter regime.

The FSP has been shown to be $NP$-hard~\cite{troyer}, and a complete solution is not in sight. However, to nevertheless obtain accurate QMC results at WDM conditions, we have introduced two novel QMC methods that are efficient at complementary parameter regimes. 
The configuration PIMC (CPIMC) method~\cite{tim_cpp15,tim3} is formulated in antisymmetric Fock-space and can be interpreted as a Monte Carlo simulation of the exact, infinite perturbation expansion around the ideal (non-interacting) system. Therefore, it excels at strong degeneracy and high density, but becomes inefficient towards strong coupling.
In contrast, the permutation blocking PIMC (PB-PIMC) approach~\cite{dornheim,dornheim2} significantly extends standard PIMC towards lower temperature and higher density, while strong coupling does not pose an obstacle. Thus, the combination of both methods allows for accurate results over a broad parameter range.

A detailed comparison of the different ranges of applicability of fermionic QMC methods at WDM conditions can be found in Ref.~\cite{dornheim_pop}.

\subsection{Dielectric approximations\label{sec:dielectric}}

The main advantage of quantum Monte Carlo methods is the exact treatment of the short-range exchange-correlation effects, which are not described accurately by any approximation.
On the other hand, the main disadvantage (despite the relatively large computational effort and non-universal range of applicability due to the sign problem) is that QMC simulations are limited to the finite simulation box. For this reason, QMC methods cannot be used to describe long-range correlations (corresponding to the limit of small wave vectors, $k\to0$). On the other hand, it has long been known that the random phase approximation (RPA) becomes exact in the limit of small $k$ for arbitrary coupling strength or temperature~\cite{kugler2}.

Furthermore, the accuracy of RPA can be significantly increased by including a so-called (static) local field correction $G(\mathbf{q})$, which is defined by the equation~\cite{kugler1}
\begin{eqnarray}\label{eq:chi}
\chi(\mathbf{q},\omega) = \frac{ \chi_0(\mathbf{q},\omega) }{ 1 - \frac{4\pi}{q^2}[1-G(\mathbf{q})]\chi_0(\mathbf{q},\omega)}\ ,
\end{eqnarray}
with $\chi(\mathbf{q},\omega)$ and $\chi_0(\mathbf{q},\omega)$ denoting the density response function of the interacting and ideal system~\cite{quantum_theory}, respectively. Furthermore, it is often convenient to compute the dielectric function
\begin{eqnarray}
\epsilon(\mathbf{k},\omega) = 1 - \frac{\chi_0(\mathbf{k},\omega)}{k^2/(4\pi)+G(\mathbf{k})\chi_0(\mathbf{k},\omega)}\ ,
\label{dielectric}
\end{eqnarray}
where the RPA limit is recovered by setting $G(\mathbf{q})=0$ in Eqs.~(\ref{eq:chi}) and (\ref{dielectric}).
Unfortunately, the local field correction is not known in practice and one has to introduce an approximation. For the UEG, the most successful approach was introduced by Singwi~\textit{et al.}~\cite{stls_original} and extended to finite temperature by Tanaka and Ichimaru~\cite{stls}. The idea is to express $G(\mathbf{q})$ as a funcional of the static structure factor
\begin{eqnarray}
G_\textnormal{STLS}(\mathbf{k}) = -\frac{1}{n} \int\frac{\textnormal{d}\mathbf{k}^\prime}{(2\pi)^3}
\frac{\mathbf{k}\cdot\mathbf{k}^\prime}{k^{\prime 2}} [S(\mathbf{k}-\mathbf{k}^\prime)-1]\; ,
\label{G}
\end{eqnarray}
which, in turn, is used again to compute the SSF via the fluctuation dissipation theorem
\begin{eqnarray}
S(\mathbf{k}) = -\frac{1}{\beta n}\sum_{l=-\infty}^{\infty} \frac{q^2}{4\pi}\left(\frac{1}{\epsilon(\mathbf{k},z_l)}-1\right)\ ,
\label{SF}
\end{eqnarray}
where the Matsubara frequencies are given by $z_l=2\pi il/\beta\hbar$.
In practice, to obtain the SSF in STLS approximation we start with 1) computing $S(k)$ in RPA, 2) use it to compute $G_\textnormal{STLS}(\mathbf{q})$ according to Eq.~(\ref{G}) and 3) subsequently obtain a new SSF from Eq.~(\ref{SF}). Steps 2) and 3) are then repeated until the structure factor and local field correction are consistent, which is the case when convergence is achieved. For completeness, we mention that first QMC results for the (static) density response function $\chi(k)$ of the warm dense electron gas have been presented in Refs.~\cite{dornheim_response,groth_response}.

\subsection{Construction of static structure factors}

\begin{figure}\center
\includegraphics[width=0.55\textwidth]{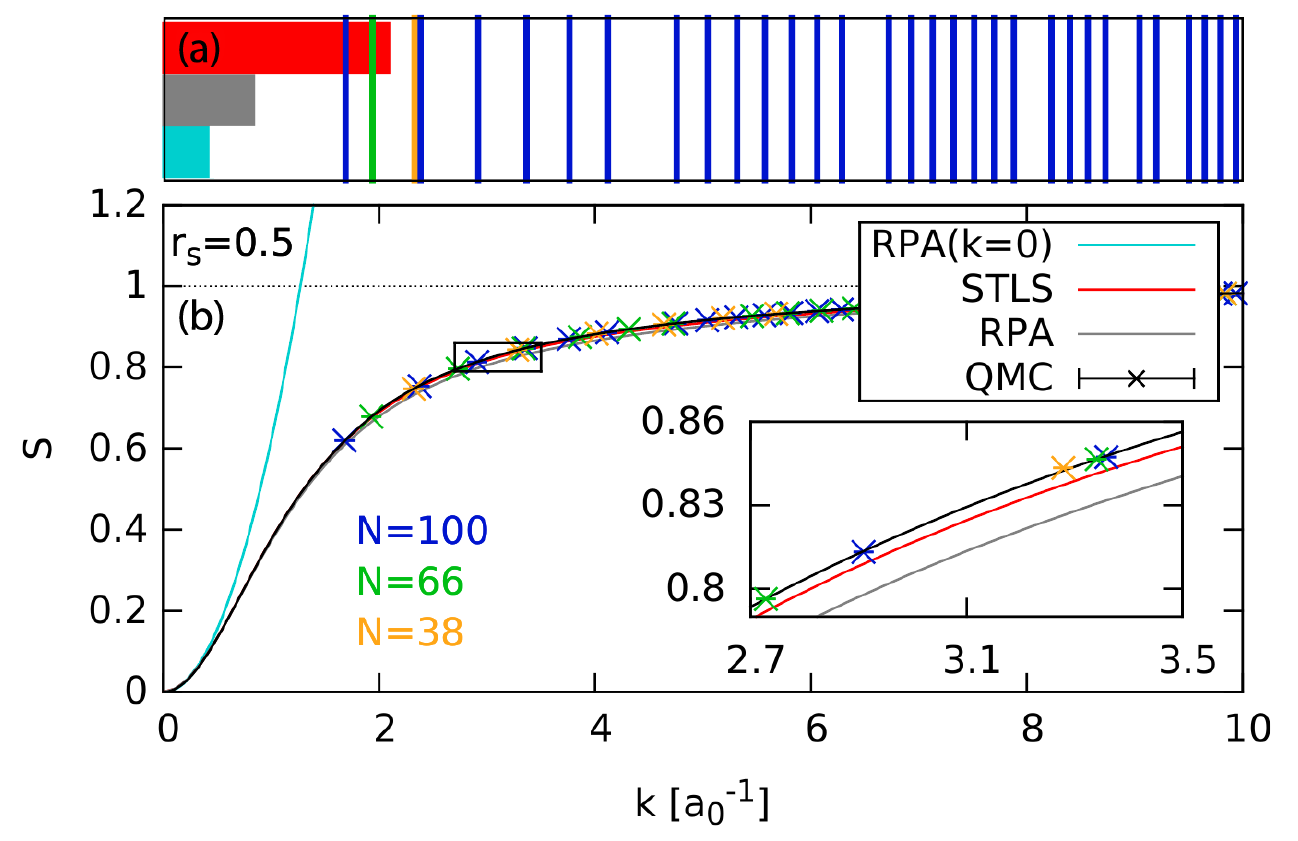}
\caption{\label{fig:prl}Schematic illustration of static structure factors for the unpolarized electron gas at $\theta=2$ and $r_s=0.5$. In panel (a), the different ranges of validity are illustrated by the light blue (RPA expansion around $k=0$, Eq.~(\ref{eq:S0})), grey (full RPA) and red (full STLS) results. The dark blue vertical lines depict the discrete $k$-grid for $N=100$ electrons. In addition, the vertical green and yellow lines show the minimum $k$-values for $N=66$ and $N=38$, respectively.
Panel (b) shows results for the static structure factor from Eq.~(\ref{eq:S0}), full RPA, full STLS, and quantum Monte Carlo (crosses) with the same three particle numbers as above. The solid black line corresponds to a spline combining STLS for small $k$ with QMC elsewhere.
Reproduced from Ref.~\cite{dornheim_prl} with the permission of the authors.
} 
\end{figure}

The construction of our new results for the static structure factor over the entire $k$-range is illustrated in Fig.~\ref{fig:prl}for the unpolarized UEG at $\theta=2$ and $r_s=0.5$. The blue vertical bars in panel (a) correspond to the discrete $k$-values (due to momentum quantization in a finite simulation cell) of a QMC simulation with $N=100$ electrons. Evidently, QMC results are not available below $k_\textnormal{min}=2\pi/L$ and the $k$-grid becomes denser for increasing $k$.
The vertical green and yellow line corresponds to the minimum $k$-value for $N=66$ and $N=38$, respectively. Furthermore, the horizontal bars illustrate the ranges of validity of an RPA expansion around $k=0$ (light blue) given by~\cite{kugler2}
\begin{eqnarray}\label{eq:S0}
S_0^\textnormal{RPA}(k) = \frac{k^2}{2\omega_p}\textnormal{coth}\left(\frac{\beta\omega_p}{2}\right)\ ,
\end{eqnarray}
the full RPA results (grey) and the full STLS data (red). For the present example, only the STLS data exhibits an overlap with the QMC results.

In panel (b), we show results for $S(k)$ itself. The crosses correspond to the QMC results for the three different particle numbers shown in panel (a). The main difference between these data sets is the different $k$-grid, while the functional form of the SSF is remarkably well converged with system size, see the inset. The light blue curve depicts the parabolic RPA expansion from Eq.~(\ref{eq:S0}), which is of interest for finite-size corrections of the interaction energy~\cite{dornheim_prl,dornheim_pop,brown}, but does not provide a sufficient description of the long-range correlations beyond the QMC data. The grey and red curves correspond to the full RPA and STLS results (see Sec.~\ref{sec:dielectric}), respectively, and are in perfect agreement with each other and Eq.~(\ref{eq:S0}) for small $k$, as expected~\cite{kugler2}. Further, the STLS curve exhibits an overlap with the QMC point at $k_\textnormal{min}$, whereas the RPA data already exhibits a minor deviation. However, for larger $k$, both STLS and RPA exhibit systematic errors, although the inclusion of the local field correction leads to a significant increase in the accuracy, see the inset.
Finally, the black line depicts a cubic basis spline (obtained using the GNU scientific library (GSL), see Ref.~\cite{gsl}) combining the red curve (for $k<k_\textnormal{min}/2$) with the blue crosses (elsewhere).
In this way, we have obtained an accurate, smooth description of the static structure factor (in the thermodynamic limit) over the entire $k$-range. All the new results presented in Sec.~\ref{sec:results} are obtained analogously.

\section{Results for the Static Structure Factor\label{sec:results}}

\begin{figure}
\includegraphics[width=0.5\textwidth]{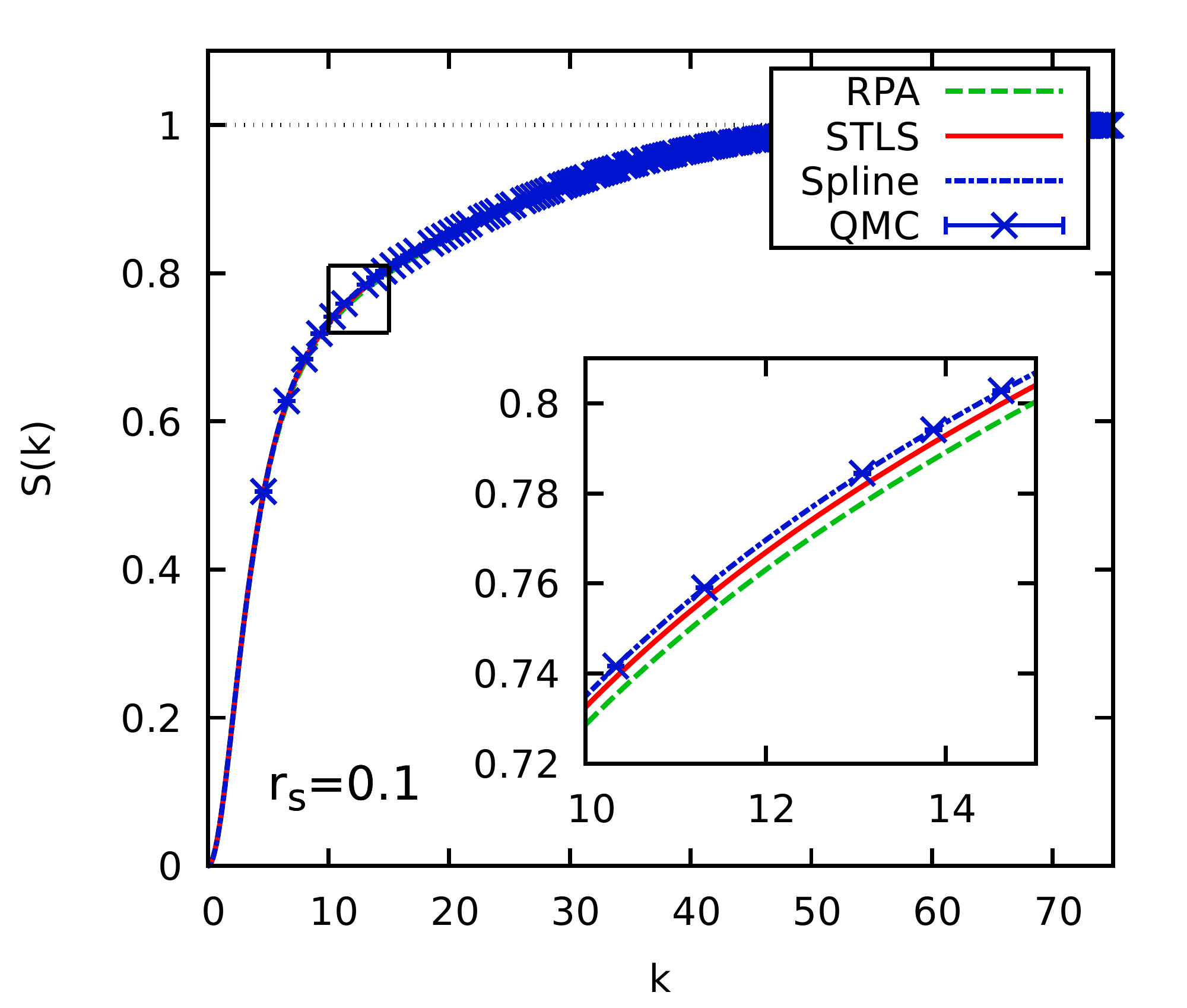}\includegraphics[width=0.5\textwidth]{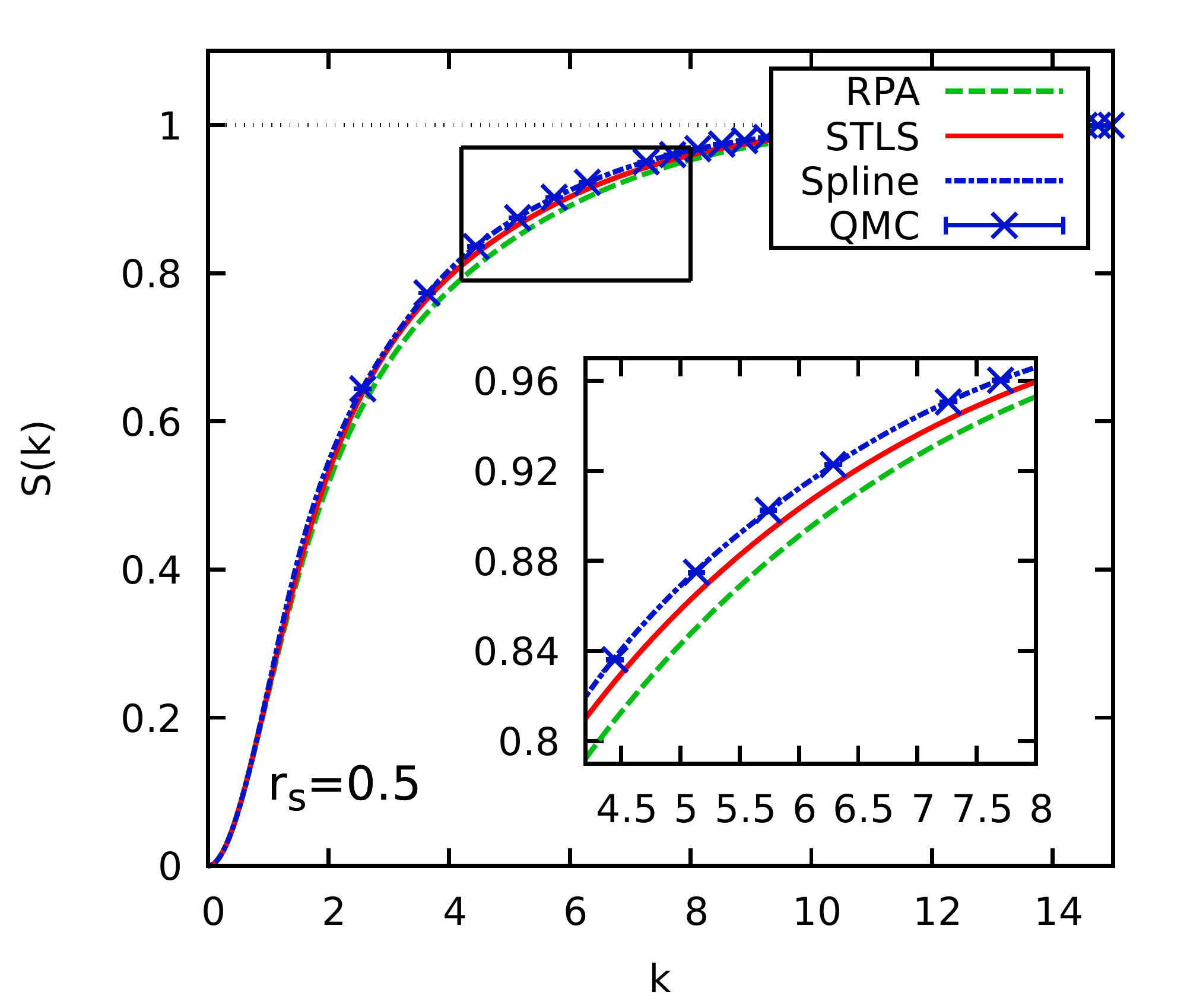}
\includegraphics[width=0.5\textwidth]{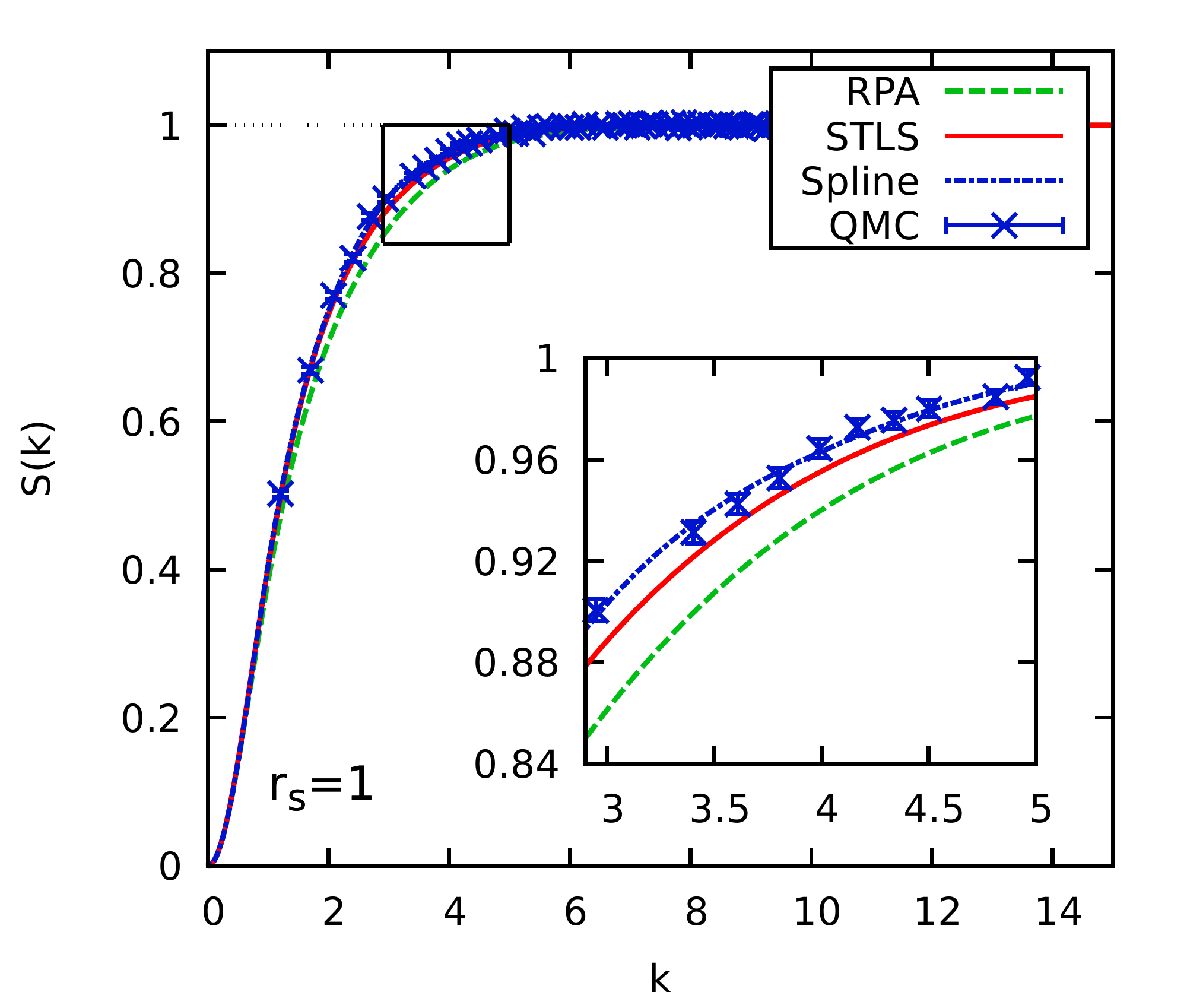}\includegraphics[width=0.5\textwidth]{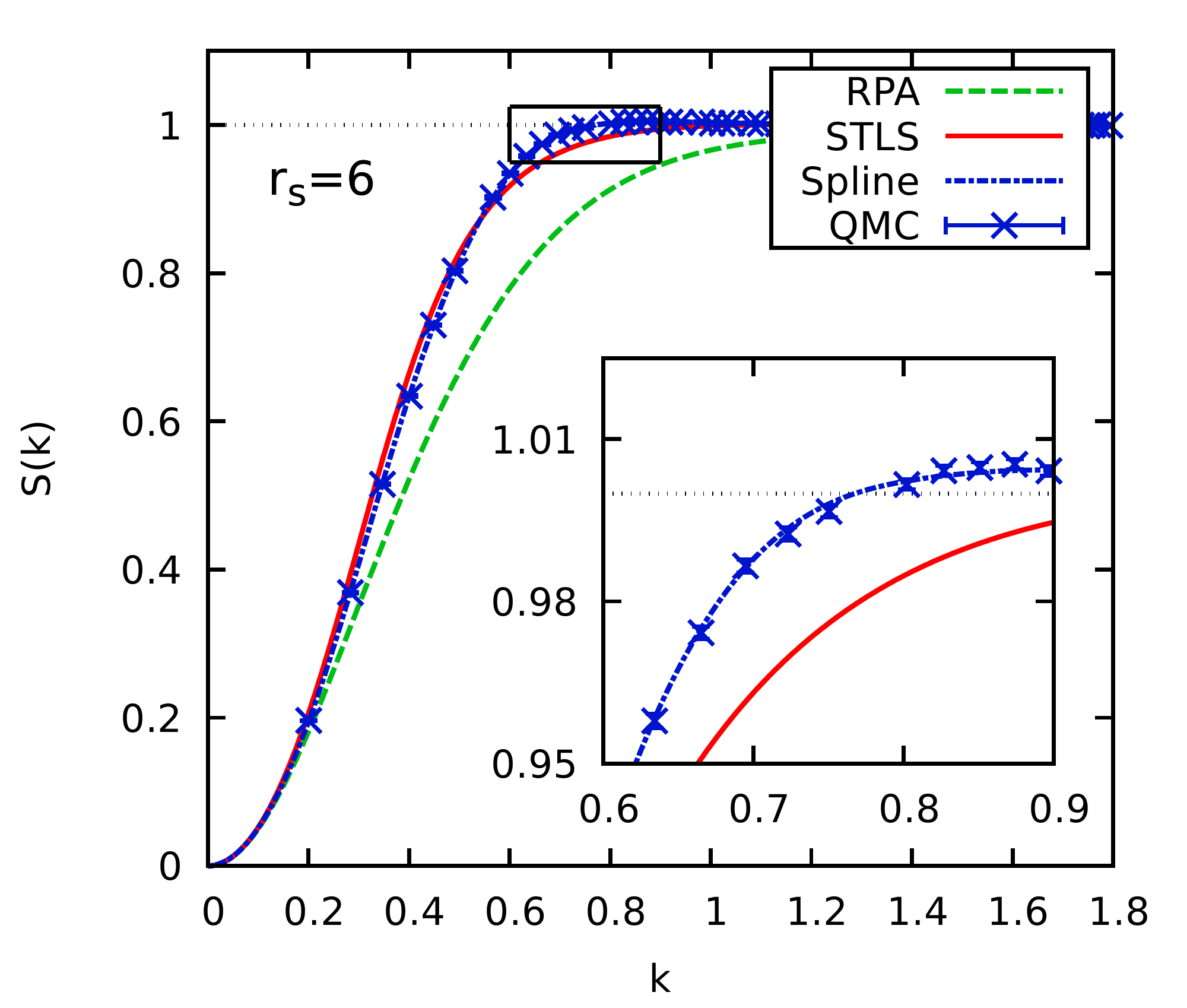}
\includegraphics[width=0.5\textwidth]{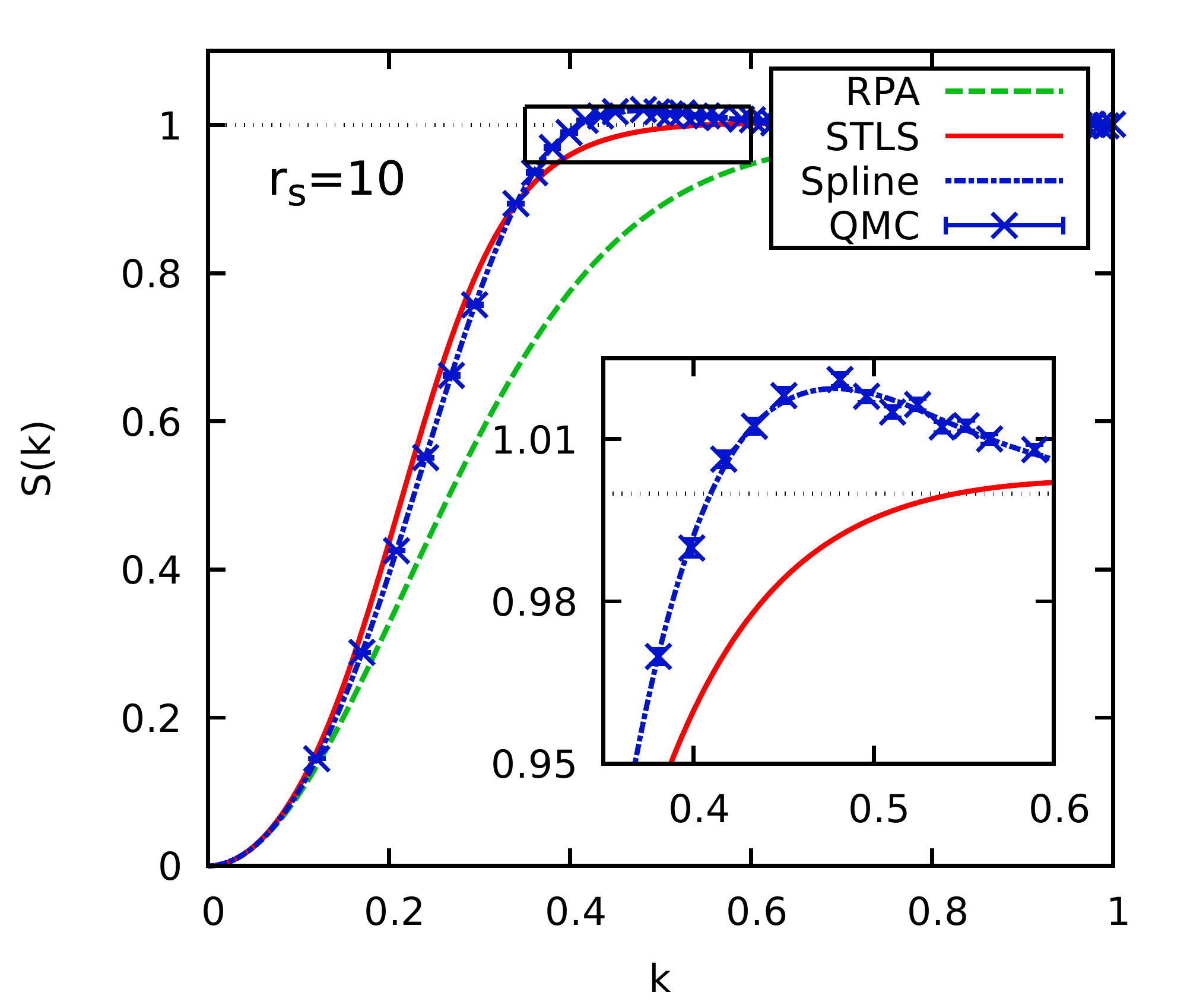}\includegraphics[width=0.5\textwidth]{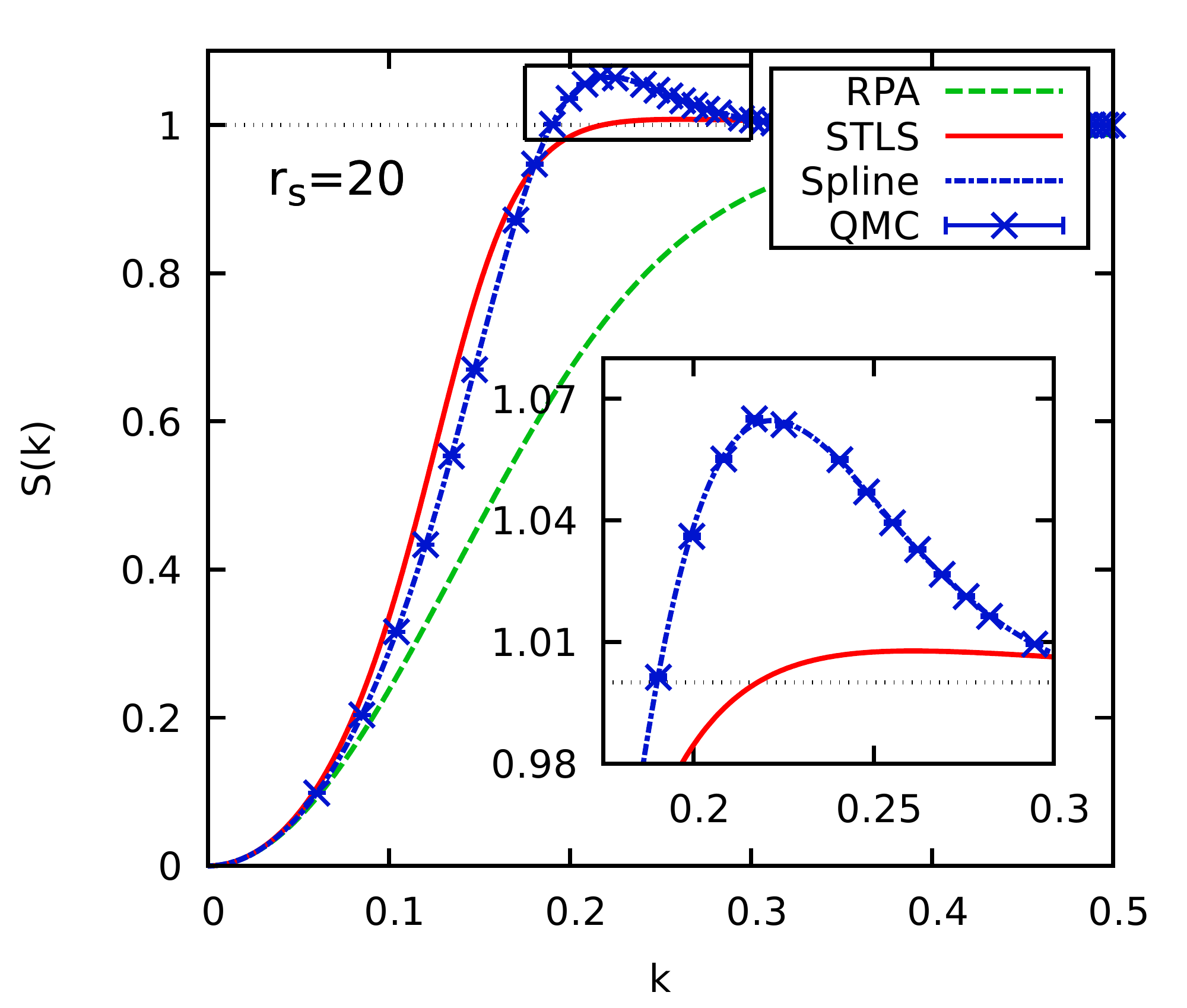}
\caption{\label{fig:rs_kachel1}Density dependence of the static structure factor at $\theta=1$ -- Shown are results for the SSF from RPA (dashed green), STLS (solid red), a cublic basis spline connecting STLS and QMC (dashed-dotted blue), and the raw QMC data (blue crosses). The depicted density parameters are $r_s=0.1,0.5,1,2,6,10,$ and $20$. All combined results for $S(k)$ are available at Ref.~\cite{github}, and selected data are given in table~\ref{table:data}.
} 
\end{figure}

Let us start our investigation with a discussion of the $r_s$-dependence of the static structure factor at $\theta=1$, which is depicted in Fig.~\ref{fig:rs_kachel1} (see also table~\ref{table:data} in the appendix).
Shown are results for the SSF from full RPA (dashed green) and STLS (solid red) calculations, quantum Monte Carlo simulations (blue crosses) and the splines connecting STLS with QMC (dash-dotted blue). For high density ($r_s=0.1$ and $r_s=0.3$), the system is only weakly non-ideal and both RPA and STLS provide an accurate description over the entire $k$-range, as it is expected. With increasing $r_s$, coupling effects become more important and especially the RPA results become substantially less accurate. In particular, the green curves are always systematically too low at intermediate $k$, which is most pronounced at $r_s=10$ and $r_s=20$, where the bias is of the order of $\Delta S/S\sim20\%$. This is due to a significant overestimation of short-range correlations, resulting in a (substantially) negative pair correlation function~\cite{hedin} at short distances.
In stark contrast, the static local field correction due to Singwi~\textit{et al.}~\cite{stls_original} significantly improves the accuracy even for large $r_s$.
Still, with increasing coupling strength there occur systematic deviations to the \textit{ab initio} QMC data. In particular, the STLS results for smaller $k$ (but not for $k\to0$, where it becomes exact) are too large, whereas they are too low in the region where $S(k)$ approaches unity. This is most evident at $r_s=20$, where the STLS approximation does not capture the maximum around $k=0.2$. Here, too, the PCF from STLS becomes negative for small $r$~\cite{stls_original}. Another fortunate feature of the STLS scheme is an error cancellation in the interaction energy per particle $v$, which can be obtained from the SSF by the relation
\begin{eqnarray}
v =
\label{EQ:v} \frac{1}{2}\int_{k<\infty}\frac{\textnormal{d}\mathbf{k}}{(2\pi)^3} \left[S(k)-1\right]\frac{4\pi}{k^2} = \frac{1}{\pi}\int_0^\infty\textnormal{d}k\ \left[S(k)-1\right]\ ,
\end{eqnarray}
where, for the second equality, we made use of the fact that the SSF only depends on the modulus of the wave vector $\mathbf{k}$ for homogeneous systems.
Therefore, the too large and too small STLS results for $S(k)$ for small and large $k$ cancel to some degree under the integral in Eq.~(\ref{EQ:v}), leading to STLS interaction energies that are more accurate than the SSF, see, e.g., Ref.~\cite{dornheim_pop}.

\begin{figure}
\includegraphics[width=0.5\textwidth]{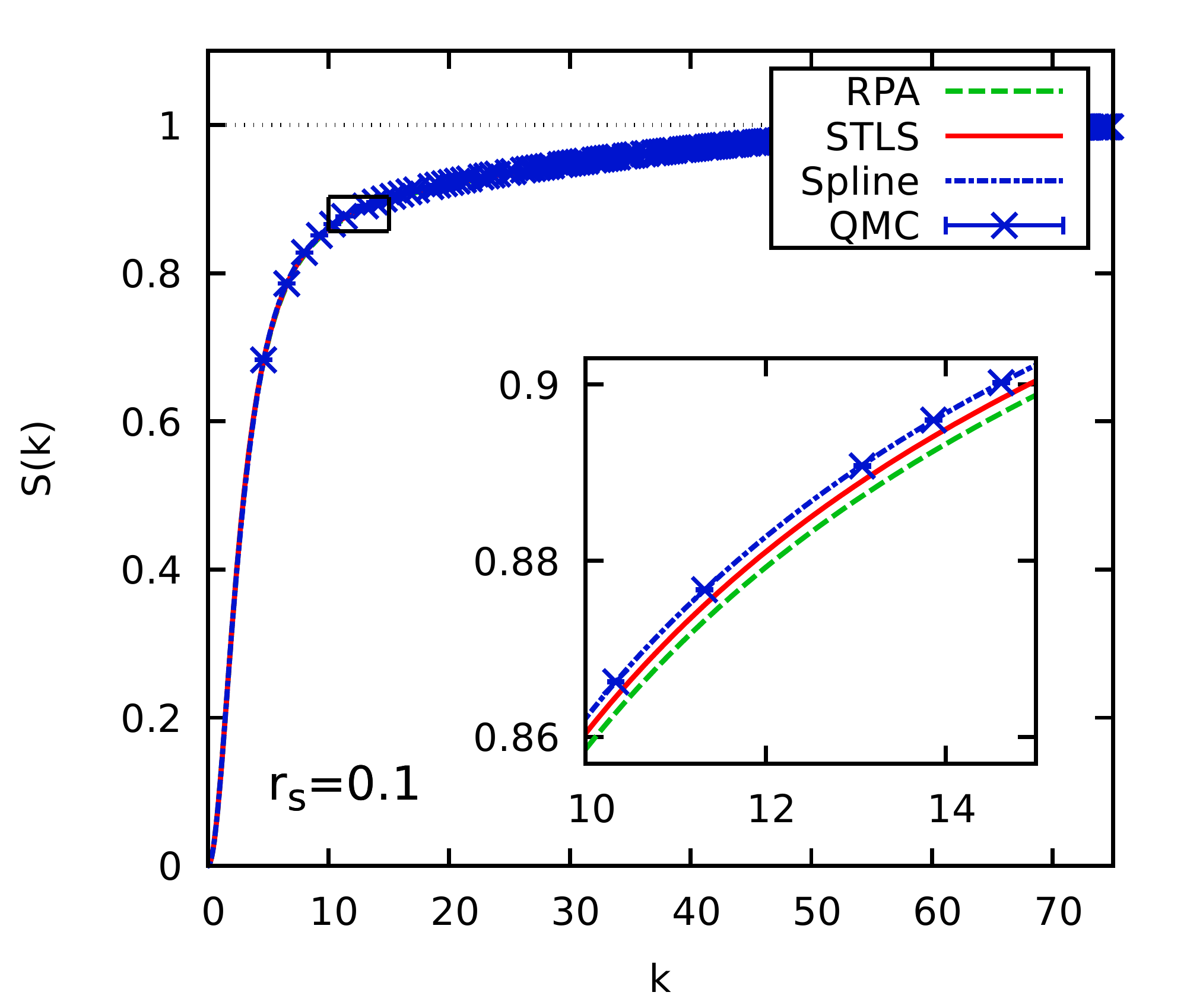}\includegraphics[width=0.5\textwidth]{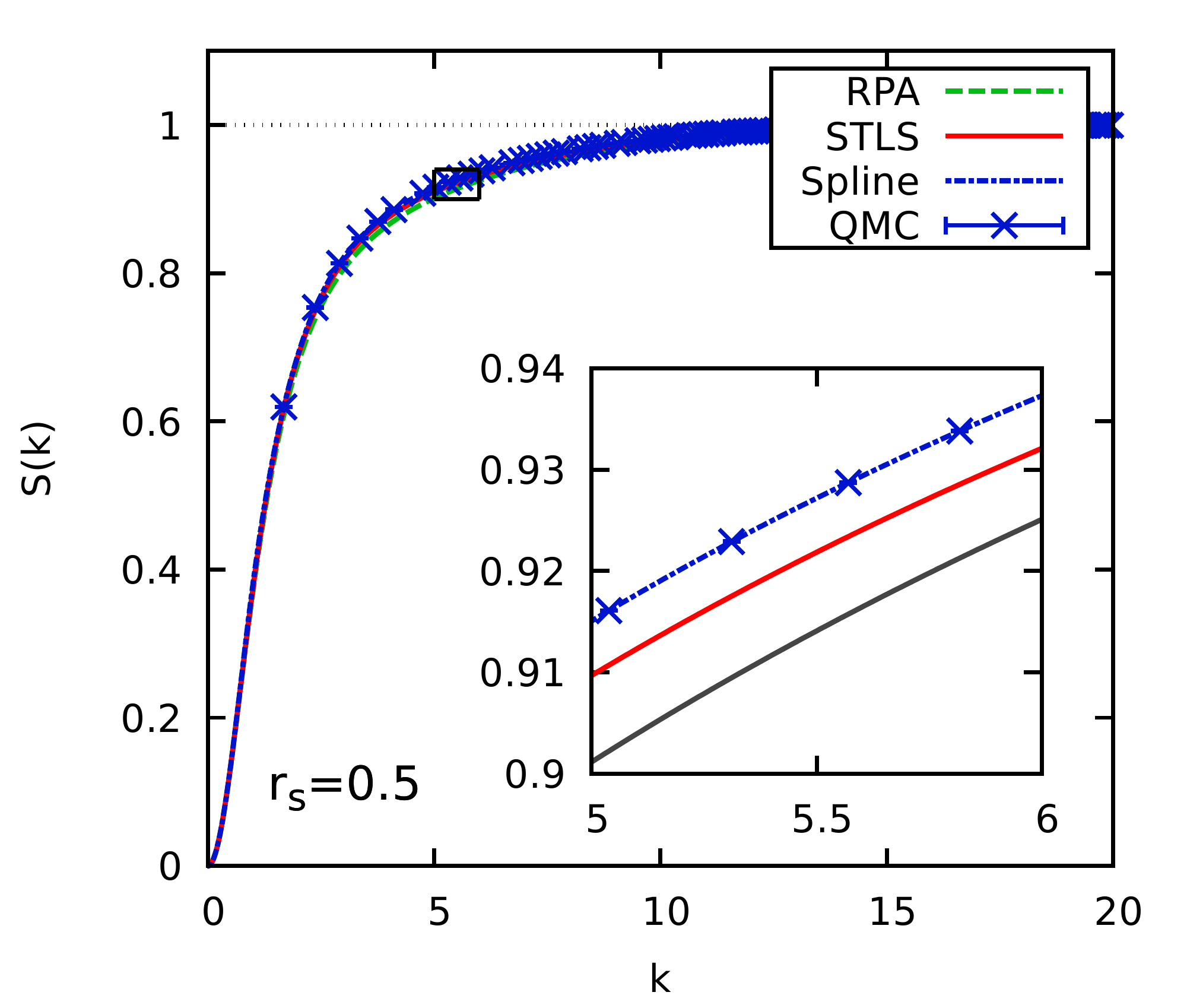}
\includegraphics[width=0.5\textwidth]{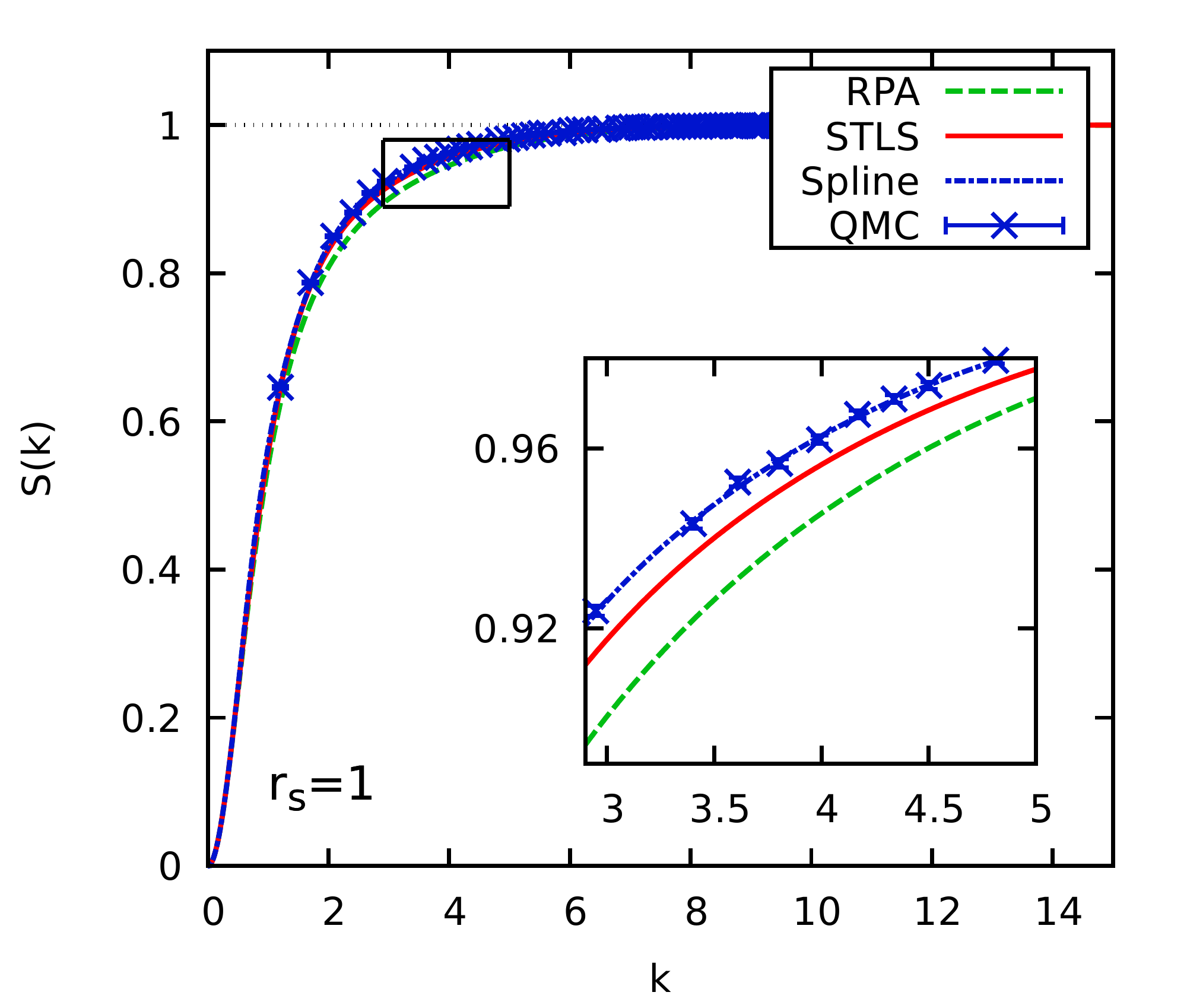}\includegraphics[width=0.5\textwidth]{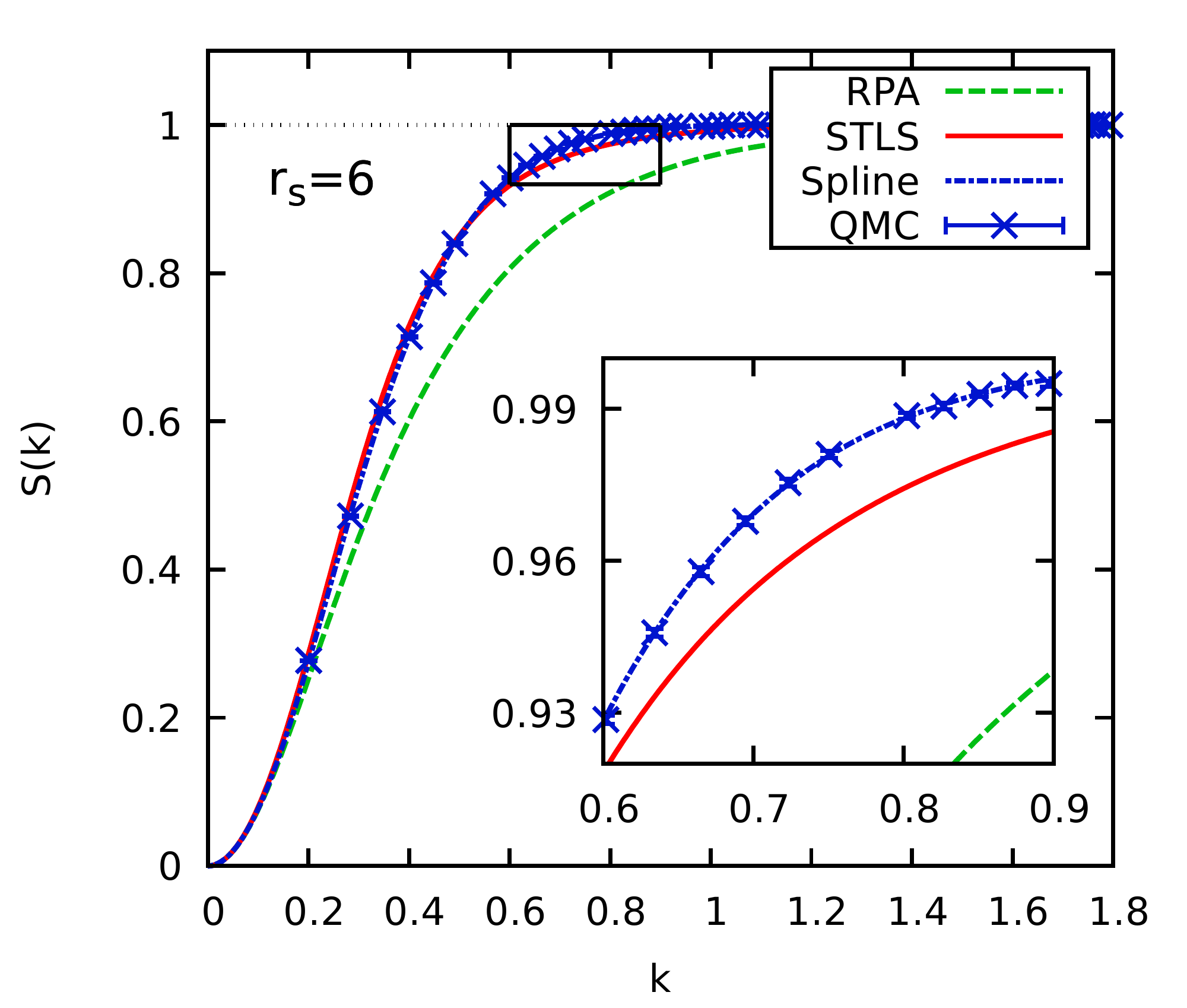}
\includegraphics[width=0.5\textwidth]{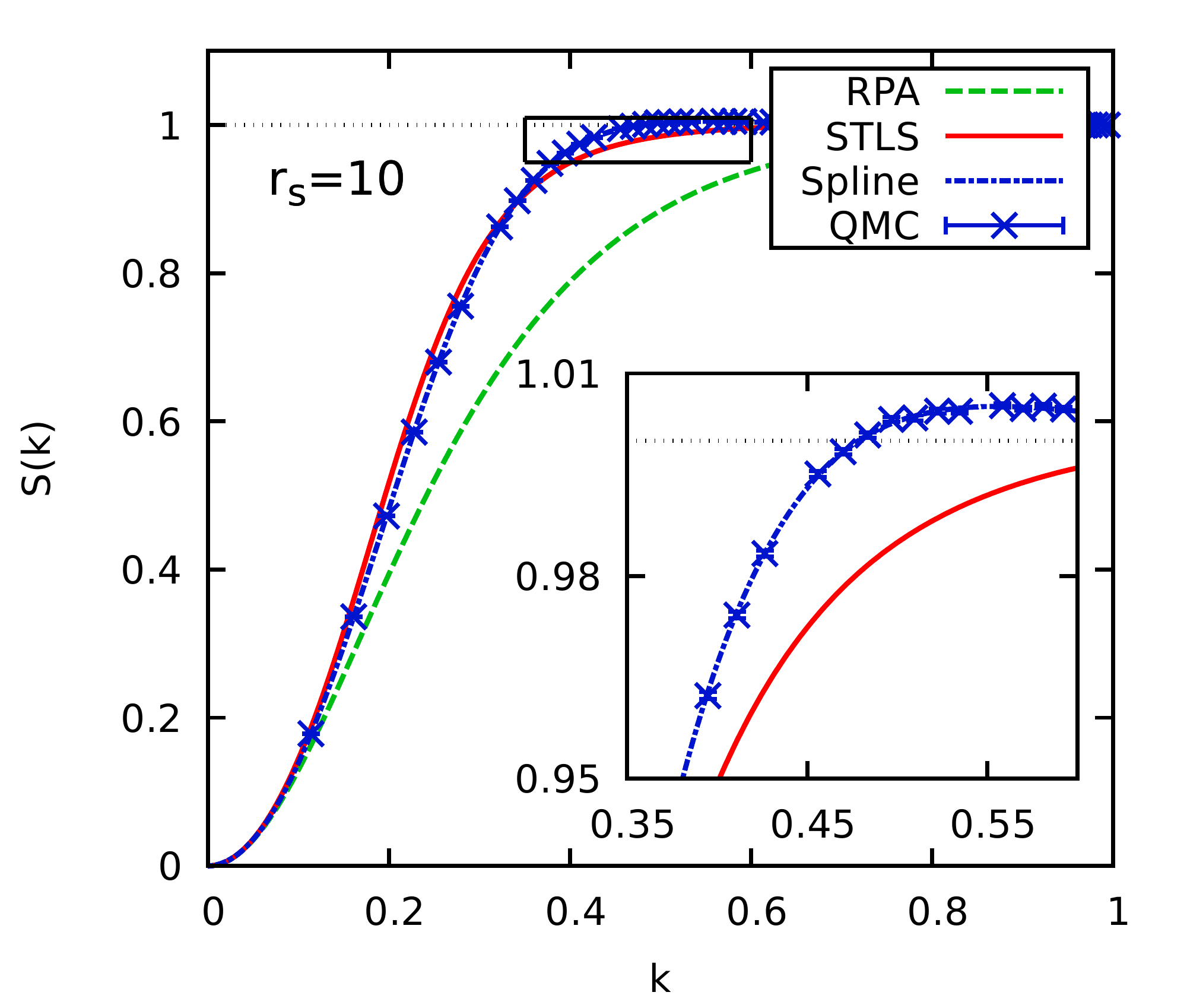}\includegraphics[width=0.5\textwidth]{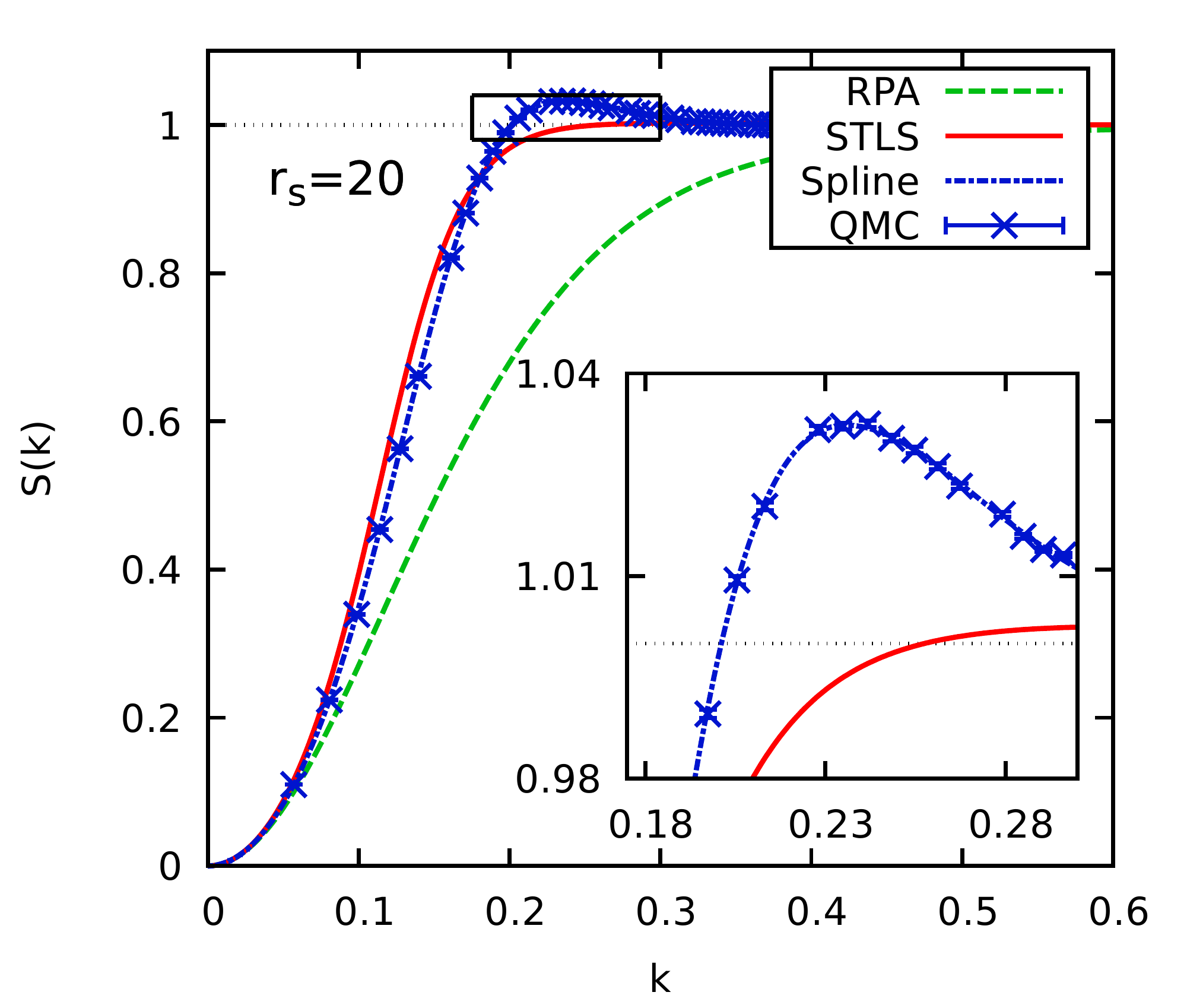}
\caption{\label{fig:rs_kachel2}Density dependence of the static structure factor at $\theta=2$ -- Shown are results for the SSF from RPA (dashed green), STLS (solid red), a cublic basis spline connecting STLS and QMC (dashed-dotted blue), and the raw QMC data (blue crosses). The depicted density parameters are $r_s=0.1,0.5,1,2,6,10,$ and $20$. All combined results for $S(k)$ are available at Ref.~\cite{github}.
} 
\end{figure}

\begin{figure}
\includegraphics[width=0.5\textwidth]{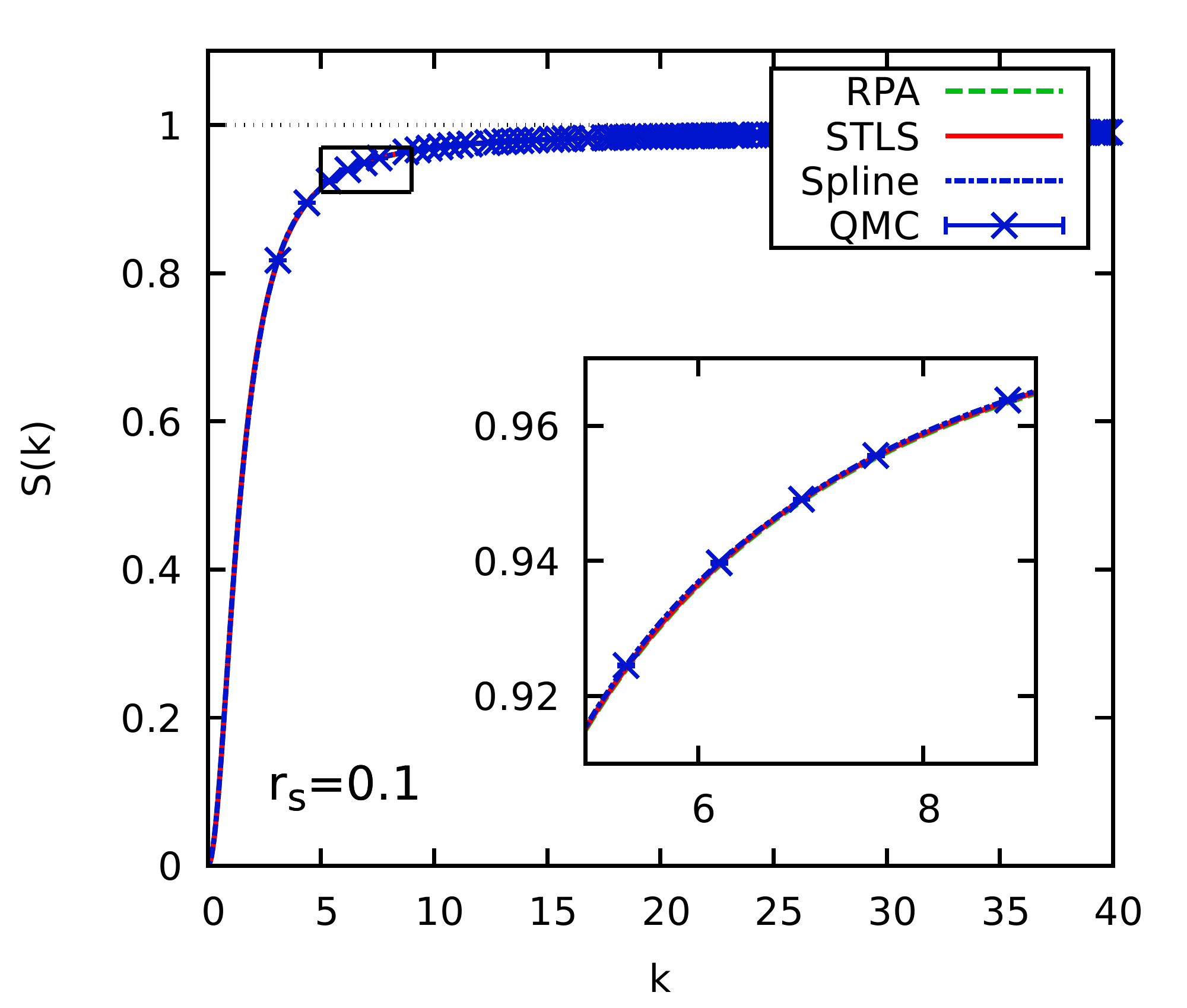}\includegraphics[width=0.5\textwidth]{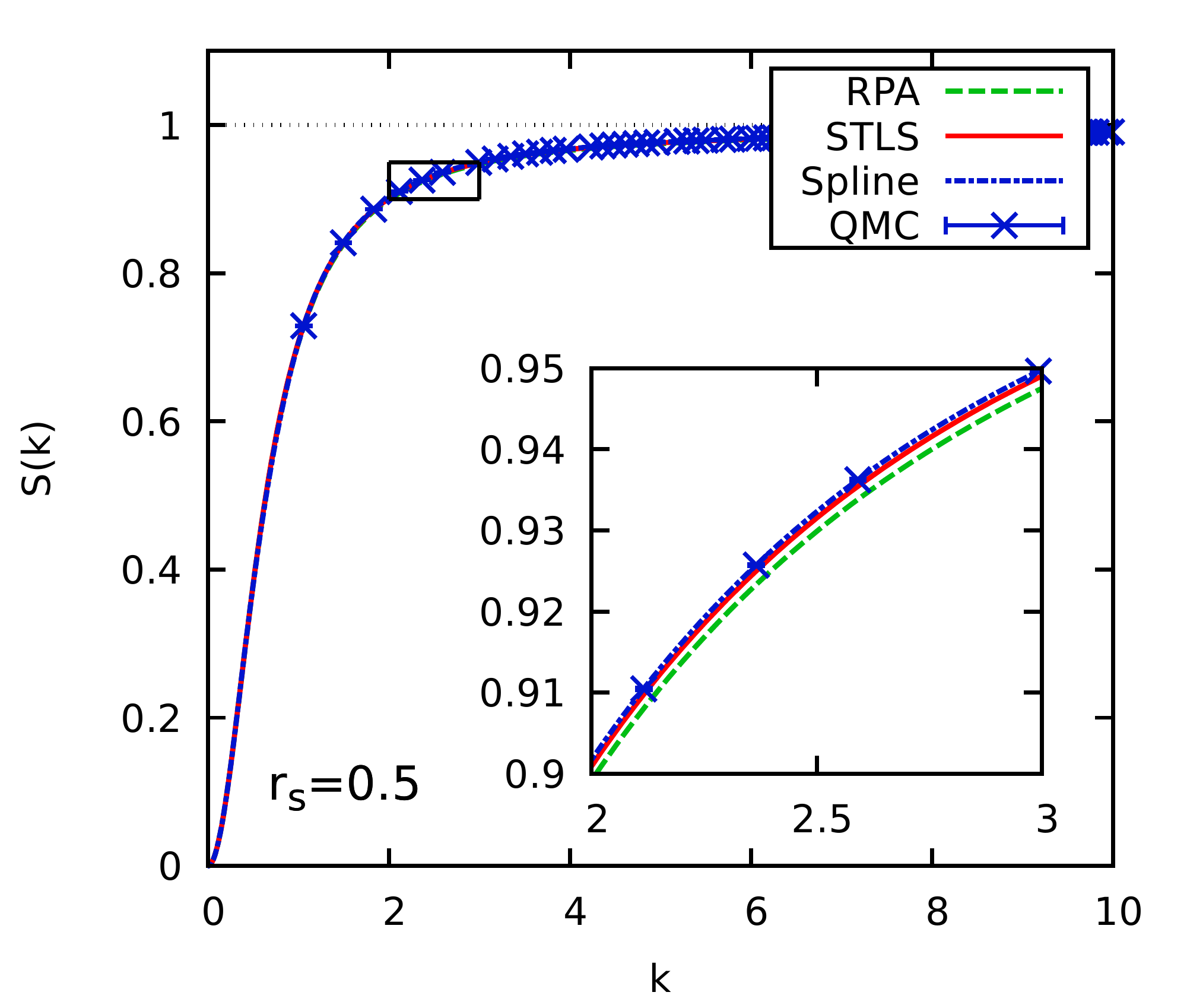}
\includegraphics[width=0.5\textwidth]{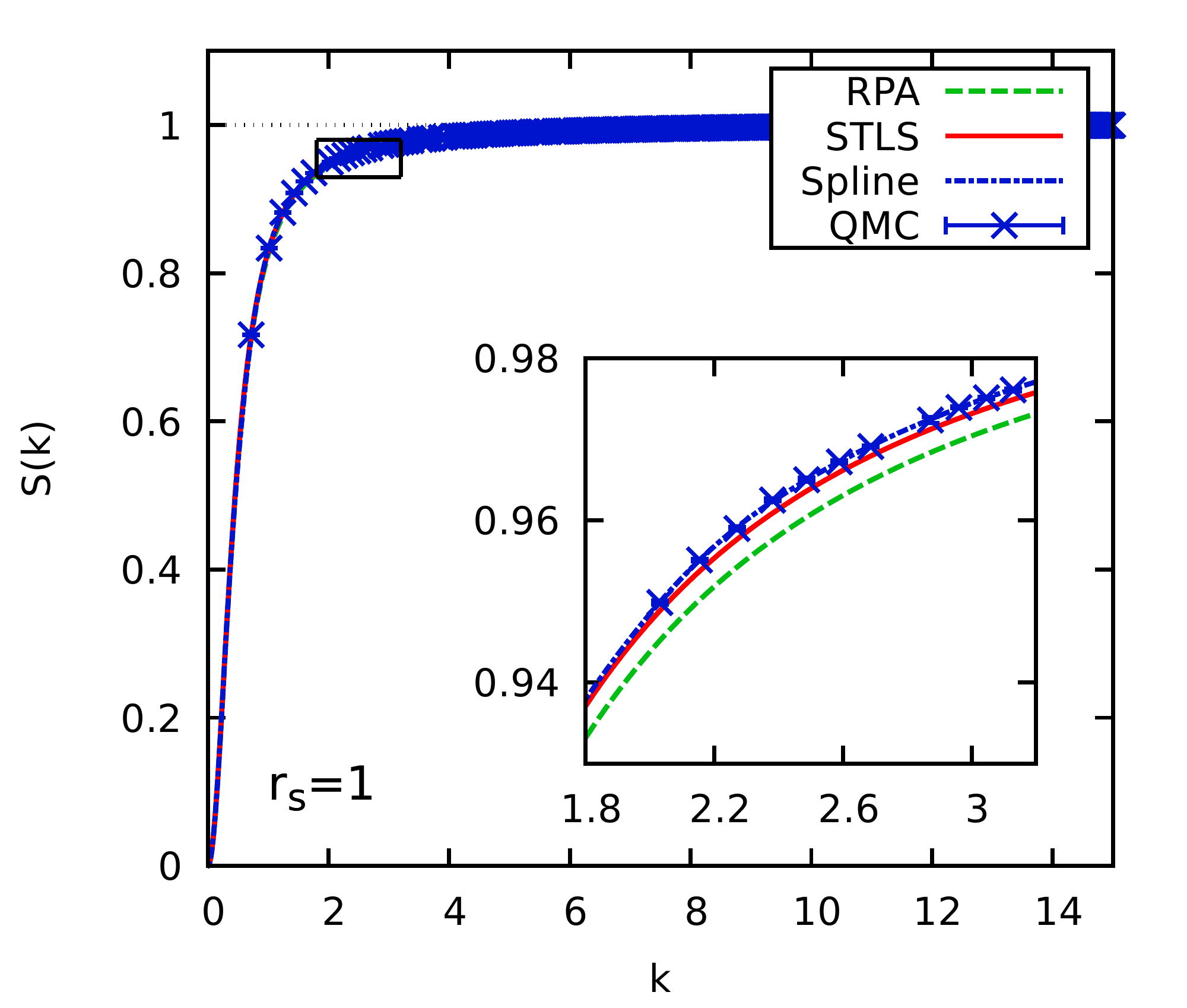}\includegraphics[width=0.5\textwidth]{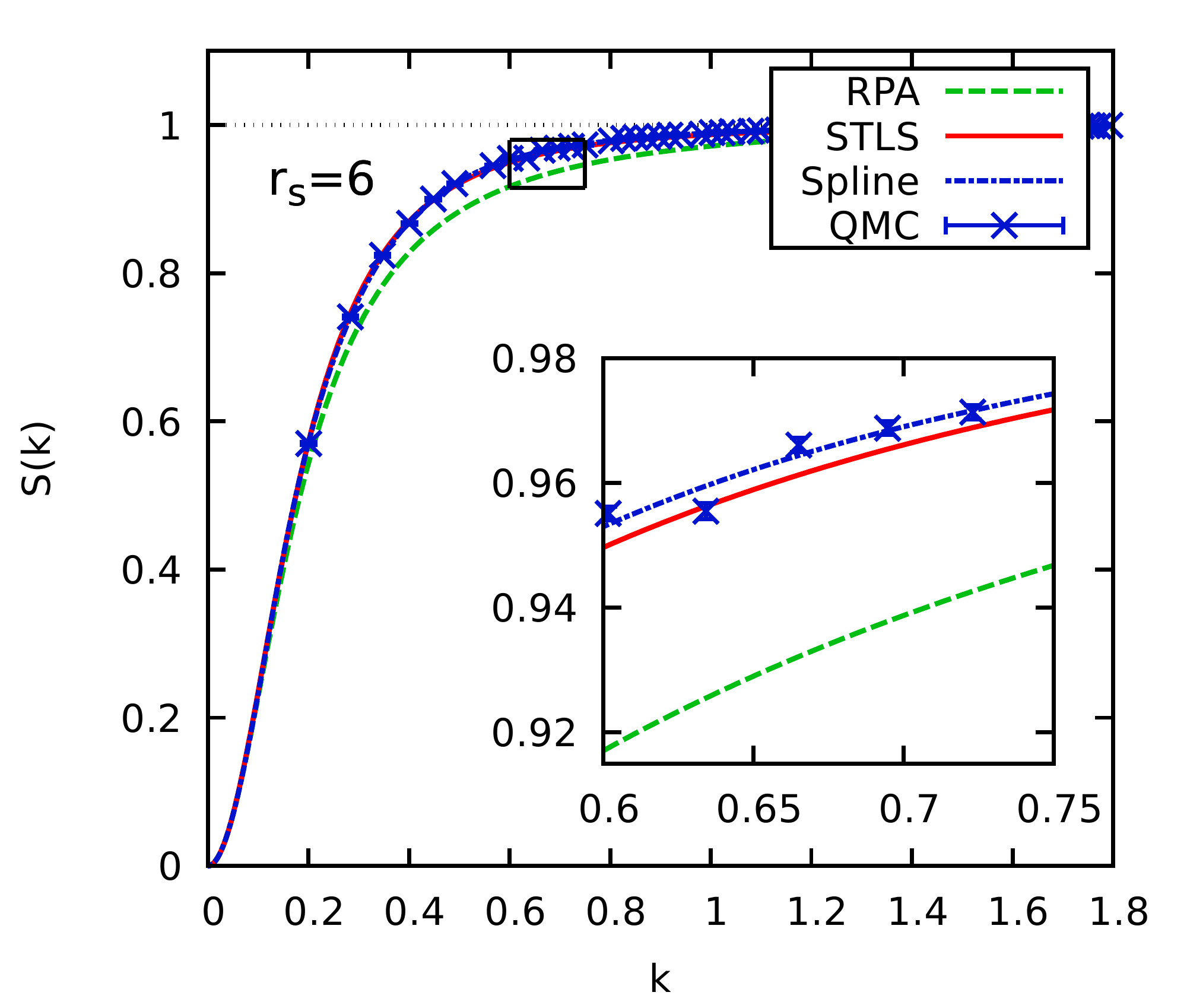}
\includegraphics[width=0.5\textwidth]{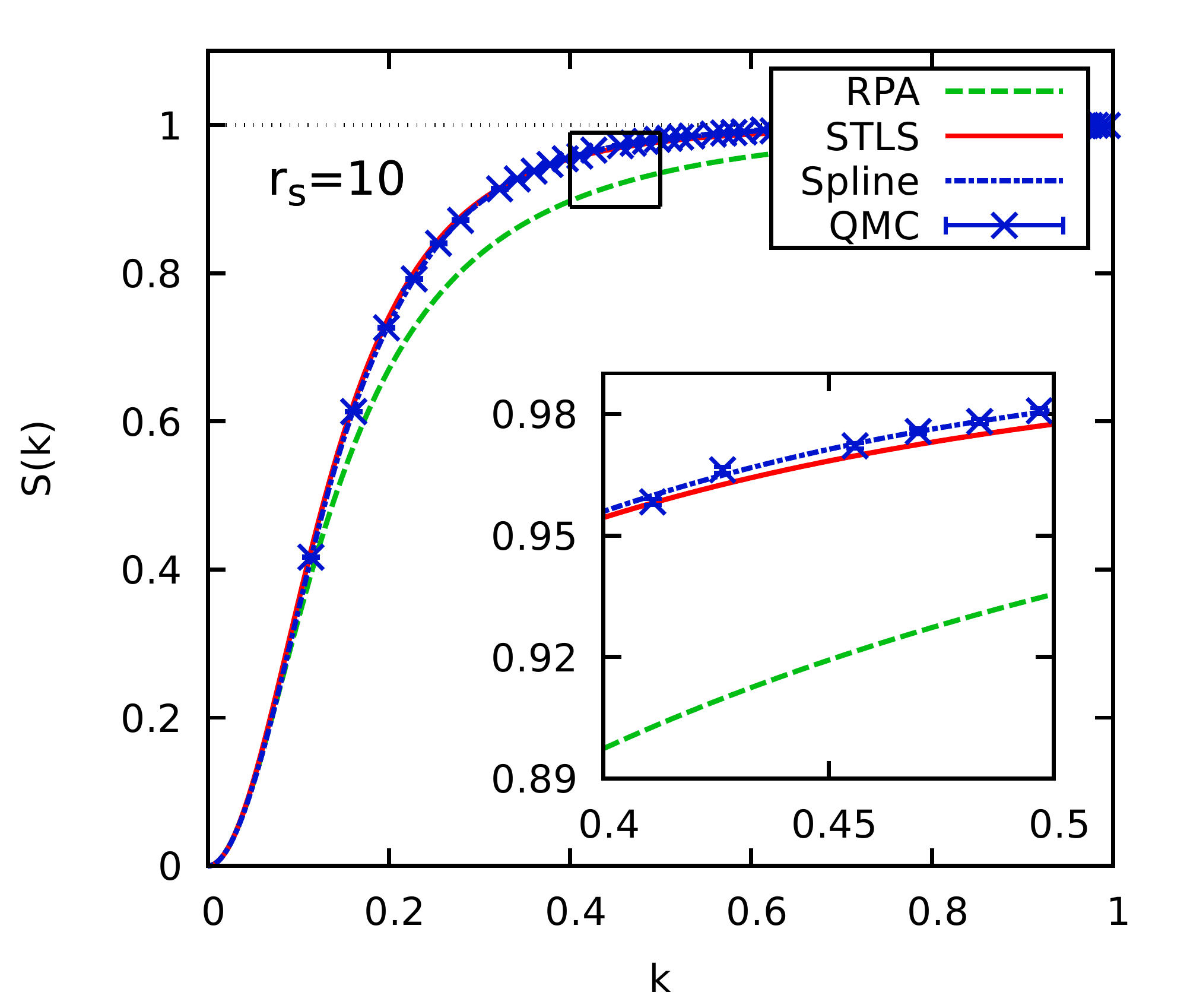}\includegraphics[width=0.5\textwidth]{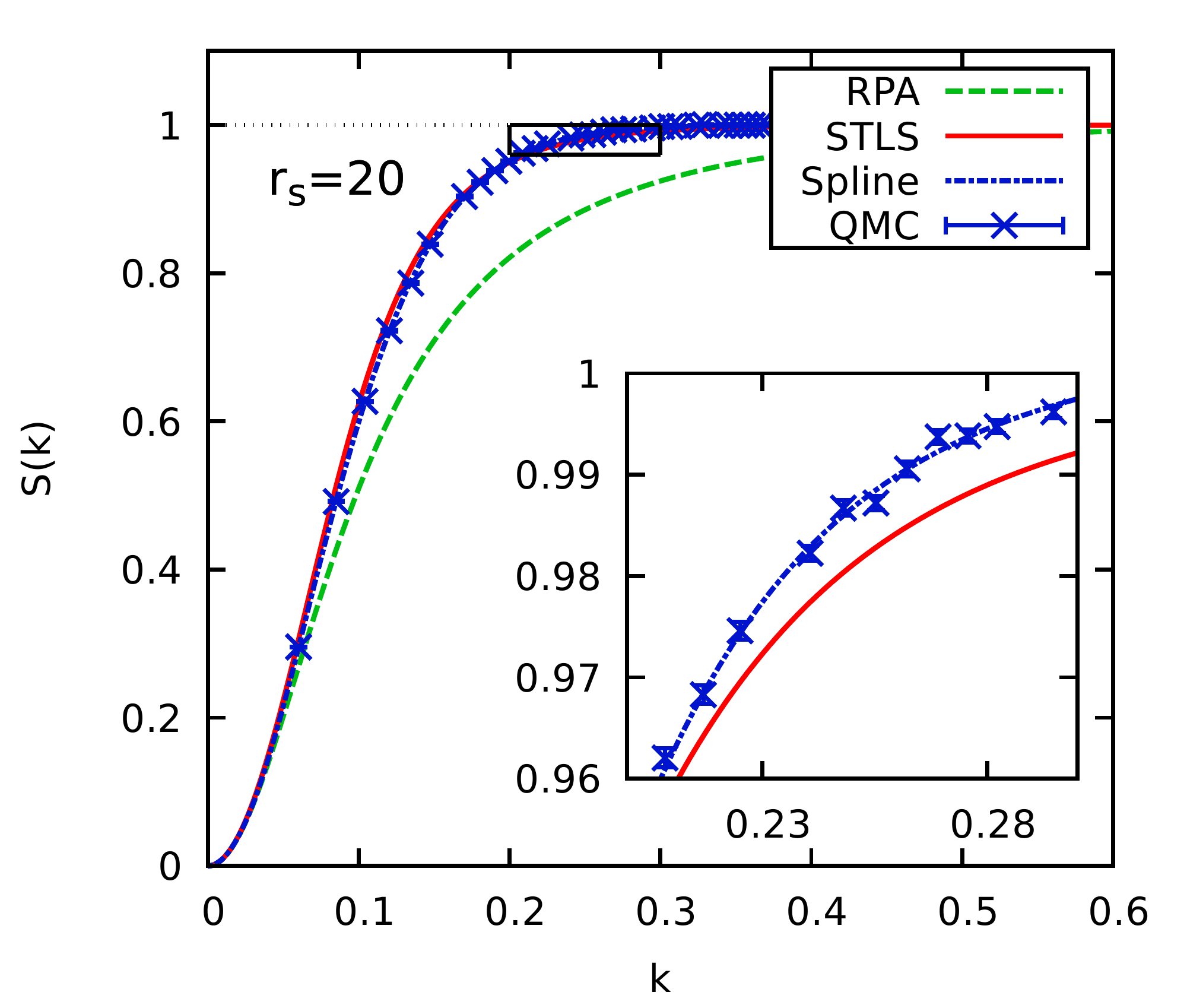}
\caption{\label{fig:rs_kachel8}Density dependence of the static structure factor at $\theta=8$ -- Shown are results for the SSF from RPA (dashed green), STLS (solid red), a cublic basis spline connecting STLS and QMC (dashed-dotted blue), and the raw QMC data (blue crosses). The depicted density parameters are $r_s=0.1,0.5,1,2,6,10,$ and $20$. All combined results for $S(k)$ are available at Ref.~\cite{github}.
} 
\end{figure}

In Figs.~\ref{fig:rs_kachel2} and \ref{fig:rs_kachel8}, we show the same information as in Fig.~\ref{fig:rs_kachel1}, but for higher temperatures, $\theta=2$ and $\theta=8$.
For $\theta=2$, the behavior of the SSF is quite similar to $\theta=1$, although the maxima at $r_s=20$ and even more so at $r_s=10$ are substantially less pronounced. 
At $\theta=8$, which corresponds to a relatively high temperature where both quantum effects and Coulomb coupling are significantly less important, the situation is quite different. In particular, the correlation-induced maximum in $S(k)$ has vanished and the STLS approximation provides an accurate description over the entire $k$-range, even for large $r_s$. The largest deviations occur at $r_s=20$, but even here $\Delta S/S$ does not exceed $1\%$. While the random phase approximation, too, becomes more accurate, there remain significant systematic errors from intermediate to large $r_s$. Therefore, we conclude that, despite the high temperature, a mean field ansatz (RPA) for the density response function, Eq.~(\ref{eq:chi}), is still not sufficient at the present parameters.

\section{Summary and Conclusion}

In summary, we have combined the exact description of the short-range exchange-correlation effects from \textit{ab initio} quantum Monte Carlo simulations with results from the Singwi-Tosi-Land-Sj\"olander (STLS) approximation, which becomes exact in the long-range limit, $k\to0$. In this way, we have been able to obtain accurate data for the static structure factor (in the thermodynamic limit) over the entire relevant $k$-range. This has allowed us to compare our new results both to the random phase approximation (RPA) and STLS over two orders of magnitude in the coupling parameter $r_s$ and for three relevant temperatures $\theta$. 
In agreement with findings in the ground state, we confirm that the RPA, due to the mean field ansatz for the density response function $\chi(\mathbf{q},\omega)$, is only accurate for weak non-ideality, but rapidly breaks down with increasing $r_s$. Even at the largest investigated temperature $\theta=8$, RPA exhibits substantial errors at intermediate $r_s$.
In stark contrast, the inclusion of the static local field correction proposed by Singwi \textit{et al.}~\cite{stls_original} significantly increases the accuracy everywhere. Only at strong coupling, $r_s=20$ and $r_s=10$, the STLS fails to accurately describe the maxima around $k=0.2$ und $k=0.5$, respectively. Furthermore, we note that due to the too large SSF for small $k$ and too small SSF for larger $k$, there occurs an error cancellation in the calculation of the interaction energy $v$, which means that STLS results for this quantity are more accurate than for $S(k)$.

We expect our new accurate static structure factors (available at Ref.~\cite{github}) of the warm dense electron gas to be of broad interest for various applications related to modern warm dense matter research. In particular, they can be used to benchmark other dielectric approximations such as quantum STLS (qSTLS)~\cite{bohm,arora} or the recent local field correction based on the hypernetted chain approximation by Tanaka~\cite{tanaka_hnc}. Furthermore, accurate data for $S(k)$ can be used to approximate the local field correction itself~\cite{hellal} or as input for the calculation of dynamic quantities using the method of frequency moments~\cite{sum}.

\begin{acknowledgement}
This work was supported by the Deutsche Forschungsgemeinschaft via project BO1366-10 and via SFB TR-24 project A9 as well as grant shp00015 for CPU time at the Norddeutscher Verbund f\"ur Hoch- und H\"ochstleistungs- rechnen (HLRN).
\end{acknowledgement}


\begin{table}
\centering
\begin{tabular}{ c c c }
  $r_s=20$ & $r_s=1$ & $r_s=0.1$ \\ \hline\hline\\ \vspace*{2.2cm}
 \begin{tabular}{c|c}
 $k$ & $S(k)$ \\   \hline \hline
 & \vspace*{-0.3cm} \\
0.00837561 & 0.00186964 \\
0.0168193 & 0.00755307 \\
0.025263 & 0.0170533 \\
0.0337067 & 0.0303731 \\
0.0421504 & 0.0475827 \\
0.0505941 & 0.0689052 \\
0.0590378 & 0.0945839 \\
0.0674815 & 0.124863 \\
0.0759252 & 0.159989 \\
0.0843689 & 0.200215 \\
0.0928126 & 0.245794 \\
0.101256 & 0.296967 \\
0.1097 & 0.353825 \\
0.118144 & 0.41634 \\
0.126587 & 0.484484 \\
0.135031 & 0.557997 \\
0.143475 & 0.634963 \\
0.151919 & 0.712752 \\
0.160362 & 0.78873 \\
0.168806 & 0.860214 \\
0.17725 & 0.924327 \\
0.185693 & 0.978152 \\
0.194137 & 1.01878 \\
0.202581 & 1.04501 \\
0.211024 & 1.05927 \\
0.219468 & 1.06441 \\
0.227912 & 1.0633 \\
0.236356 & 1.05815 \\
0.244799 & 1.05047 \\
0.253243 & 1.04169 \\
0.261687 & 1.03321 \\
0.27013 & 1.02564 \\
0.278574 & 1.01905 \\
0.287018 & 1.01352 \\
0.295461 & 1.00907 \\
0.303905 & 1.00562 \\
0.312349 & 1.00302 \\
0.320793 & 1.00115 \\
0.329236 & 0.999848 \\
0.33768 & 0.999023
 \end{tabular}

  & 
  
  \begin{tabular}{c|c}
   $k$ & $S(k)$ \\   \hline \hline
    & \vspace*{-0.3cm} \\
0.147084 & 0.014099 \\
0.29553 & 0.0539904 \\
0.443975 & 0.113561 \\
0.592421 & 0.18666 \\
0.740867 & 0.267137 \\
0.889313 & 0.348842 \\
1.03776 & 0.425663 \\
1.1862 & 0.494009 \\
1.33465 & 0.554356 \\
1.4831 & 0.60755 \\
1.63154 & 0.654436 \\
1.77999 & 0.695858 \\
1.92843 & 0.732662 \\
2.07688 & 0.765664 \\
2.22532 & 0.795312 \\
2.37377 & 0.821801 \\
2.52222 & 0.845323 \\
2.67066 & 0.86607 \\
2.81911 & 0.884233 \\
2.96755 & 0.900005 \\
3.116 & 0.913588 \\
3.26444 & 0.925229 \\
3.41289 & 0.935189 \\
3.56134 & 0.943728 \\
3.70978 & 0.951108 \\
3.85823 & 0.957587 \\
4.00667 & 0.963426 \\
4.15512 & 0.968789 \\
4.30356 & 0.973683 \\
4.45201 & 0.978098 \\
4.60046 & 0.982026 \\
4.7489 & 0.985456 \\
4.89735 & 0.98838 \\
5.04579 & 0.990793 \\
5.19424 & 0.99273 \\
5.34268 & 0.994262 \\
5.49113 & 0.995456 \\
5.63958 & 0.996383 \\
5.78802 & 0.99711 \\
5.93647 & 0.997708
  \end{tabular}

  & 
  
  \begin{tabular}{c|c}
   $k$ & $S(k)$ \\   \hline \hline
    & \vspace*{-0.3cm} \\
1.46222 & 0.11347 \\
2.94033 & 0.322402 \\
4.41844 & 0.484183 \\
5.89656 & 0.593229 \\
7.37467 & 0.662812 \\
8.85278 & 0.70891 \\
10.3309 & 0.741495 \\
11.809 & 0.766743 \\
13.2871 & 0.787148 \\
14.7652 & 0.804333 \\
16.2433 & 0.819583 \\
17.7215 & 0.833513 \\
19.1996 & 0.846342 \\
20.6777 & 0.858384 \\
22.1558 & 0.869857 \\
23.6339 & 0.880723 \\
25.112 & 0.891012 \\
26.5901 & 0.900838 \\
28.0682 & 0.910186 \\
29.5464 & 0.918993 \\
31.0245 & 0.927267 \\
32.5026 & 0.935014 \\
33.9807 & 0.94223 \\
35.4588 & 0.94892 \\
36.9369 & 0.955095 \\
38.415 & 0.960744 \\
39.8931 & 0.965865 \\
41.3713 & 0.970485 \\
42.8494 & 0.974635 \\
44.3275 & 0.978345 \\
45.8056 & 0.981625 \\
47.2837 & 0.984483 \\
48.7618 & 0.986971 \\
50.2399 & 0.989136 \\
51.718 & 0.991 \\
53.1962 & 0.992587 \\
54.6743 & 0.993928 \\
56.1524 & 0.995055 \\
57.6305 & 0.995999 \\
59.1086 & 0.99678
  \end{tabular} 
  
  \vspace*{-2cm} \\
  \hline\hline
\end{tabular}
\caption{Static structure factor $S(k)$ for the unpolarized electron gas at $\theta=1$ (see Fig.~\ref{fig:rs_kachel1}) -- all data have been obtained by combining STLS data for small $k$ with QMC data elsewhere. Extensive data for $\theta=1,2,4,8$ and multiple $r_s$ values are availabe at Ref.~\cite{github}. }
\label{table:data}
\end{table}

\end{document}